\documentclass[journal]{IEEEtran}
\usepackage{upgreek}
\usepackage{graphicx}
\usepackage{subfigure}
\usepackage{amsmath,bm}
\usepackage{mathrsfs,amsmath}
\usepackage{enumerate}
\usepackage[mathscr]{euscript}
\usepackage[square, comma, sort&compress, numbers]{natbib}
\usepackage{color}
\usepackage{url}

\ifCLASSINFOpdf
\else
\fi

\begin{document}


\title{\begin{Huge}Transmission Model for Resonant Beam SWIPT with Telescope Internal Modulator\end{Huge}}

\author{Wen Fang,
Yunfeng Bai,
Qingwen Liu,~\IEEEmembership{Senior Member,~IEEE},
Shengli Zhou,~\IEEEmembership{Fellow,~IEEE},

\thanks{
W.~Fang, Y.~Bai, and Q.~Liu are with the College of Electronic and Information Engineering, Tongji University, Shanghai 200000, China (e-mail: wen.fang@tongji.edu.cn, baiyf@tongji.edu.cn, and qliu@tongji.edu.cn).
}

\thanks{S.~Zhou is with the Department of Electrical and Computer Engineering, University of Connecticut, Storrs, CT 06250, USA (e-mail: shengli.zhou@uconn.edu).}

}

\maketitle

\begin{abstract}

\normalsize
To satisfy the long-range and energy self-sustaining communication needs of electronic devices in the Internet of Things (IoT), we introduce a simultaneous wireless information and power transfer (SWIPT) system using the resonant beam that incorporates a telescope modulator inside a cavity for suppressing diffraction losses. We theoretically analyze power transfer in the resonant beam system with telescope internal modulator (TIM-RBS) considering the electromagnetic field propagation, the end-to-end (E$2$E) power transfer, and power and information reception. The numerical evaluation demonstrates that the TIM can effectively compress the beam spot, which allows the TIM-RBS to transmit energy twice as far as the RBS without TIM at higher power. Additionally, the largest transmission distance and maximum output power are proportional to the input power, and about $34\rm{m}$ transmission distance, $4\rm{W}$ electric power, and $12\rm{bps/Hz}$ spectral efficiency can be achieved in the TIM-RBS with $200\rm{W}$ input power. Hence, TIM-RBS can be considered as a promising option for realizing long-range, high-power, and high-rate SWIPT.

\end{abstract}


\IEEEpeerreviewmaketitle

\section{Introduction}\label{Section1}
With the rapid development of the Internet of Things (IoT) and $5$th Generation Mobile Communication ($5$G) technologies, the power supply and high-rate communication for electronic devices (e.g., sensors, cameras) have become well-known issues \cite{wu2014cognitive, sun2019cooperative}. Recently, to support novel applications in the IoT and 5G wireless networks, some studies have been conducted to implement self-sustainable communication systems with energy harvesting (EH) techniques \cite{perera2018simultaneous}.

\begin{figure}[!t]
    \centering
    \includegraphics[scale=0.55]{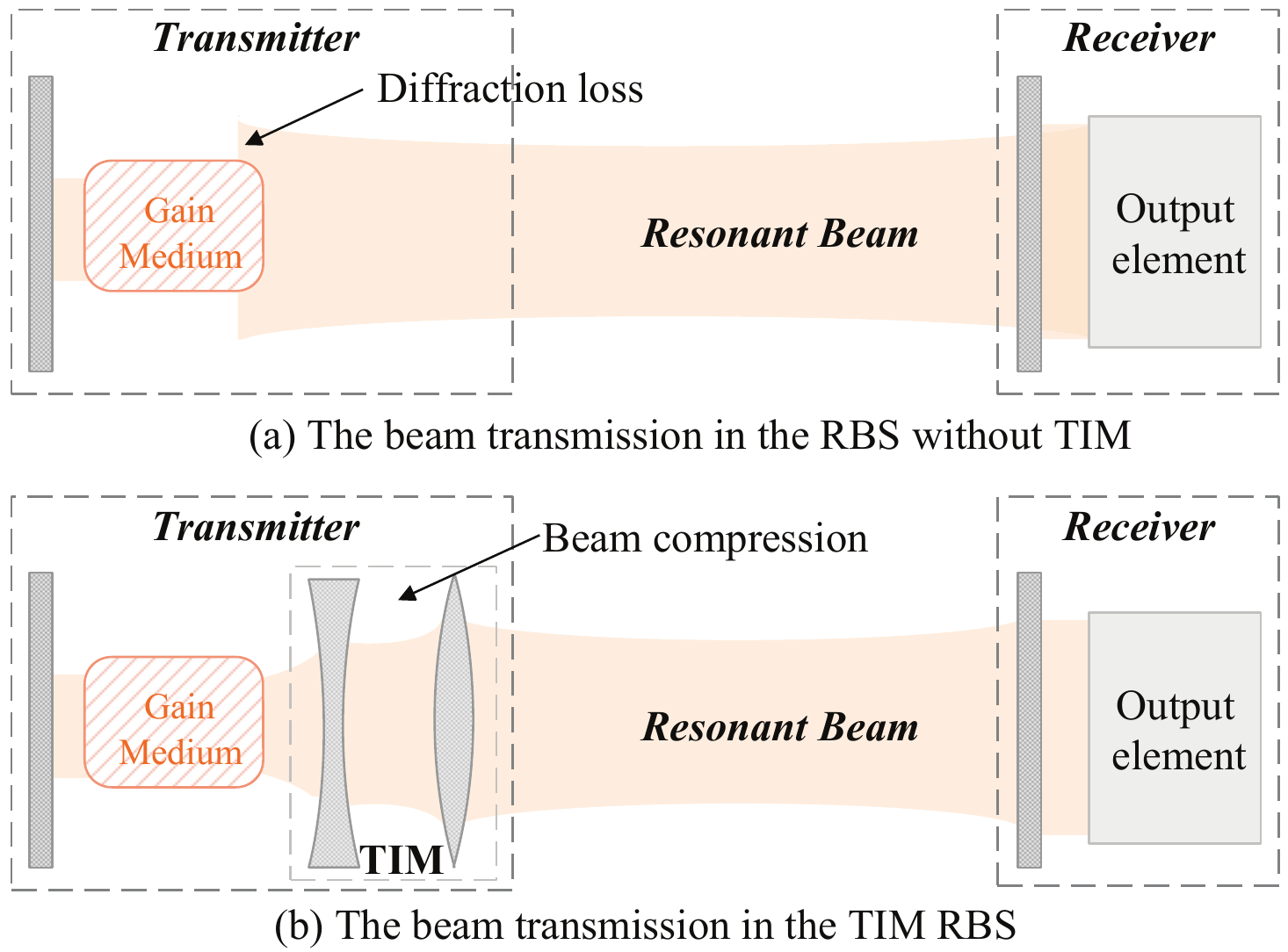}
    \caption{The comparison of the beam transmission in the RBS without TIM and TIM-RBS.}
    \label{Fig1}
\end{figure}

Compared with the currently available batteries and charged super-capacitors, EH can capture and convert wasted or important energy into electricity, which makes it possible to integrate energy sources into wireless networks \cite{hou2016preliminary, niyato2007wireless}. However, since the irregular and unforeseeable nature of ambient sources, the application of EH faces many limitations and applies only to specific environments \cite{krikidis2014simultaneous}.

The wireless power transfer (WPT) technology is one of the novel EH technologies that can overcome the above limitations \cite{lu2015wireless}. Recent WPT technologies that have been widely used include near-field and far-field energy transmission \cite{garnica2013wireless}. For near-field WPT, such as magnetic induction and magnetic resonance, watt-level power can be transferred within multi-centimeter-level distances using centimeter-sized transceivers \cite{jawad2017opportunities}. For far-field WPT, radio-frequency (RF) or laser can transfer milliwatt-level energy over meter-level distances under safety regulations \cite{xia2015efficiency}. 
The prospect of integrating WPT into communication networks creates a demand for technology that can simultaneously transmit information and power to electronic devices \cite{krikidis2014simultaneous}. Thus, the concept of simultaneous wireless information and power transfer (SWIPT) was first introduced in \cite{varshney2008transporting} with theoretical analysis.

To realize long-range, high-power WPT and high-rate communication, the existing near-field and far-field SWIPT face the challenge of balancing distance, power, spectral efficiency, and safety \cite{ulukus2015energy, chen2016secrecy, ku2015advances}. The latest SWIPT, resonant beam system (RBS), is firstly proposed in \cite{fang2021end}, where the end-to-end power transfer was theoretically and numerically studied, and the $5\rm{m}$ distance was exhibited. The resonant beam (i.e. the intra-cavity laser) is used to transmit energy between the spatially separated transmitter and receiver \cite{liu2016dlc, zhang2018distributed2}. However, due to the resonant beam spot diffuses with the increase of transmission distance, the transmission loss grows with the extension of cavity length between the transmitter and the receiver. Afterwards, if the transmission distance keeps extending, the resonance condition in the RBS will be no longer satisfied (i.e. losses are greater than gain), so the energy transfer will be suspended, which will result in limited transmission distances \cite{wang2018channel}.

In this paper, we introduce an RBS with a telescope internal modulator (TIM), which can be called TIM-RBS. As shown in Fig.~\ref{Fig1}(b), the telescope contains a concave mirror and a convex mirror, which has been used in the laser resonator for obtaining reliable operation of an Nd:YAG laser with a large-volume TEM$_{00}$ mode in \cite{hanna1981telescopic}. Then, due to the telescope can reduce the size of the beam to increase the diffraction per unit length, Sarkies \cite{sarkies1979stable} reported using a telescope in an Nd:YAG resonator allowing easily controllable adjustment to compensate thermal lensing under varied pumping conditions. In Fig.~\ref{Fig1}(b), the resonant beam carrying energy is transmitted from the transmitter to the receiver through input reflector, gain medium, TIM, and output reflector in order while it passes through input reflector, gain medium, and output reflector in the RBS without TIM depicted in Fig.~\ref{Fig1}(a). From \cite{hanna1981telescopic, sarkies1979stable} and Fig.~\ref{Fig1}, the main contribution of TIM is the compression of resonant beam spot, which brings small transmission losses and allows more beam energy to enter the gain medium for gain amplification, releasing more energy to be transmitted to the receiver for charging and communication. Thus, the transmission distance will increase with a higher power in the TIM-RBS. 

\begin{figure}[!t]
    \centering
    \includegraphics[scale=0.54]{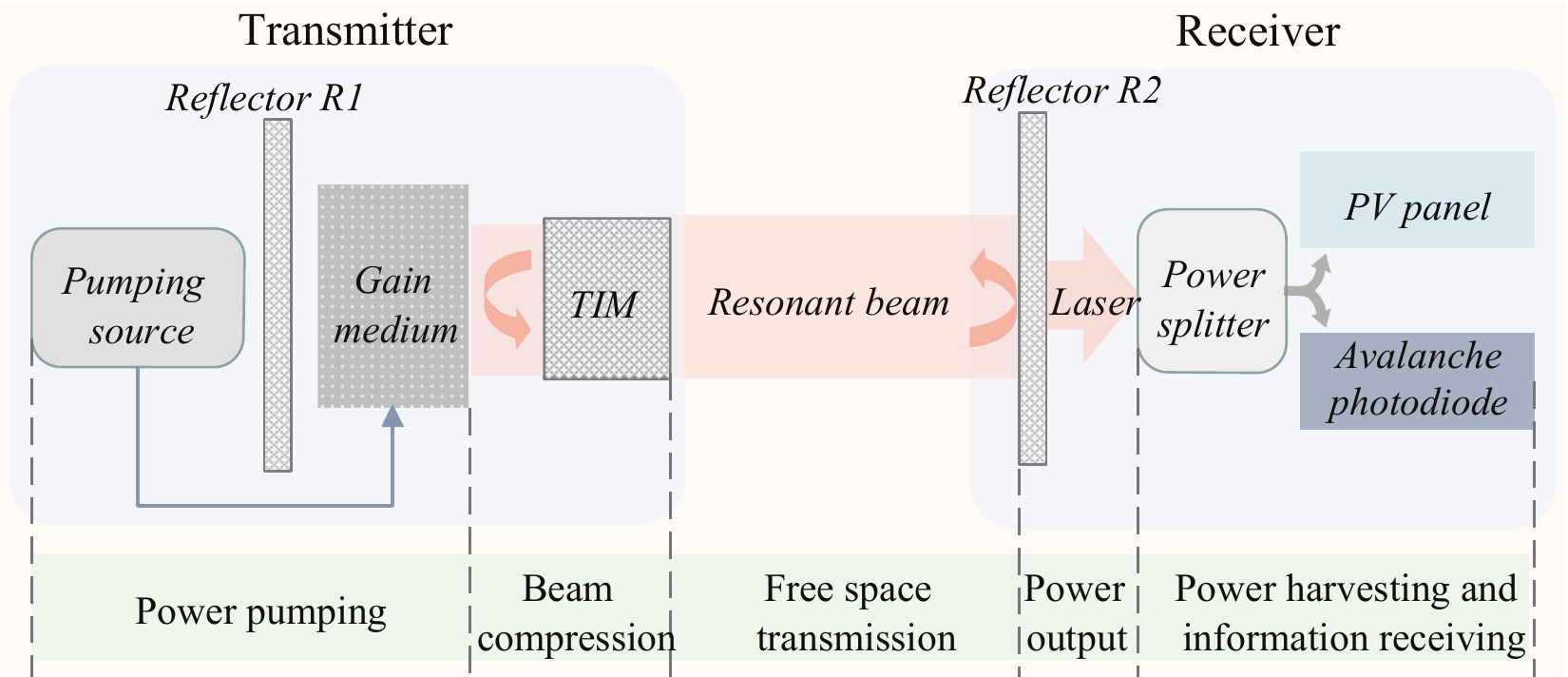}
    \caption{The structure and E$2$E power transmission model of the TIM-RBS.}
    \label{Fig2}
\end{figure}

The contributions of this paper include:
\begin{enumerate}
    \item [\bf c1)] We propose an accurate transmission model for the TIM-RBS, which combines the electromagnetic field propagation and E$2$E power transfer model and allows to obtain precise energy distribution on the optical plane, transmission efficiencies, and output power.
    \item [\bf c2)] Using the proposed accurate transmission model, we visualise the process of beam spot compression and verify that the TIM-RBS can achieve longer range and higher power compared with the RBS without TIM.
    \item [\bf c3)] The numerical evaluation shows that the TIM-RBS can realize SWIPT at a distance of two times longer than the RBS without TIM, and about $34\rm{m}$ transmission distance, $4\rm{W}$ electric power, and $12\rm{bps/Hz}$ spectral efficiency can be achieved in the TIM-RBS with $200\rm{W}$ input power.
\end{enumerate}

In the rest of this paper, the structure of the TIM-RBS, the effect of the TIM, and the E$2$E power transfer will be illustrated in Section II, the accurate transmission model including electromagnetic field propagation, E$2$E power transfer model, power and information reception will be proposed in Section III. Then, Section IV will depict the performance comparison of TIM-RBS and RBS without TIM and the power transfer evaluation of the TIM-RBS. Finally, the conclusion with future research of the TIM-RBS will be presented in Section V.

\section{System Overview}\label{Section2}
To realize long-range power and information transfer using resonant beam, the telescope internal modulator (TIM) is adopted in the RBS compressing the beam spot for decreasing transmission loss. In this section, we will illustrate the structure of the TIM-RBS and propose the power transfer model.

\subsection{TIM-RBS System}\label{}
The TIM-RBS shown in Fig.~\ref{Fig2} incorporates spatially separated transmitter and receiver structures, with the lasing gain medium situated in the transmitter and the transmitter's output directed into the receiver.

The implementation of a TIM-RBS comprises:

i) a transmitter containing:
a) a pumping source providing the input power for the system;
b) an input reflector R$1$ having high reflectivity, usually 100\rm{\%};
c) a gain medium offering sites for photon emission;
d) a TIM compressing the resonant beam for decreasing the diffraction loss.

ii) a receiver including:
a) an output reflector R$2$ with partial reflectivity;
b) a power splitter (PS) dividing the received power into two power streams based on a certain PS ratio;
c) a photovoltaic (PV) panel converting the received beam power into electric power for charging the battery;
d) an avalanche photodiode (APD) converting the received beam power to electronic signals for communication.

\begin{figure}[!t]
    \centering
    \includegraphics[scale=0.58]{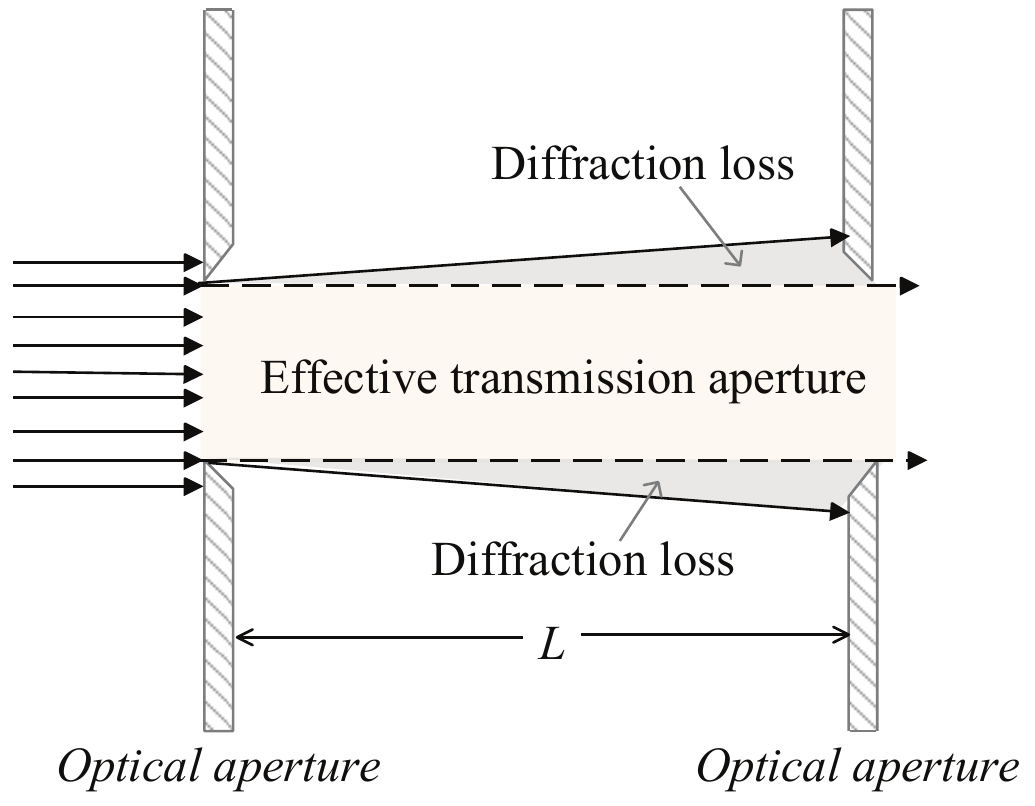}
    \caption{The diffraction loss due to optical apertures.}
    \label{Fig_diffractionloss}
\end{figure}

For a TIM-RBS, the transmitter can be located in a fixed position and properly sealed from dust and other contaminants, while the receiver is embedded in electronic devices for charging and communication. The resonant beam carrying the power transmits in the free space between the transmitter and the receiver. During the power transmission in the system, the resonant beam will pass through a series of optical apertures, e.g. input/output reflectors, gain medium. Since the limitations of optical aperture size, the transmission undergoes a series of diffraction losses. That is, in a resonant beam system, some part of the resonant beam will be lost either by spillover at the reflectors or by limiting apertures, such as the lateral boundaries of the gain medium. As shown in Fig.~\ref{Fig_diffractionloss}, these losses depend on the diameter of the beam in the plane of the aperture and the aperture radius, the diffracted beam energy outside the effective aperture of the optical components (reflectors or gain medium) will be lost, thus generating diffraction loss \cite{li1965diffraction}.

\begin{figure}[!t]
    \centering
    \includegraphics[scale=0.58]{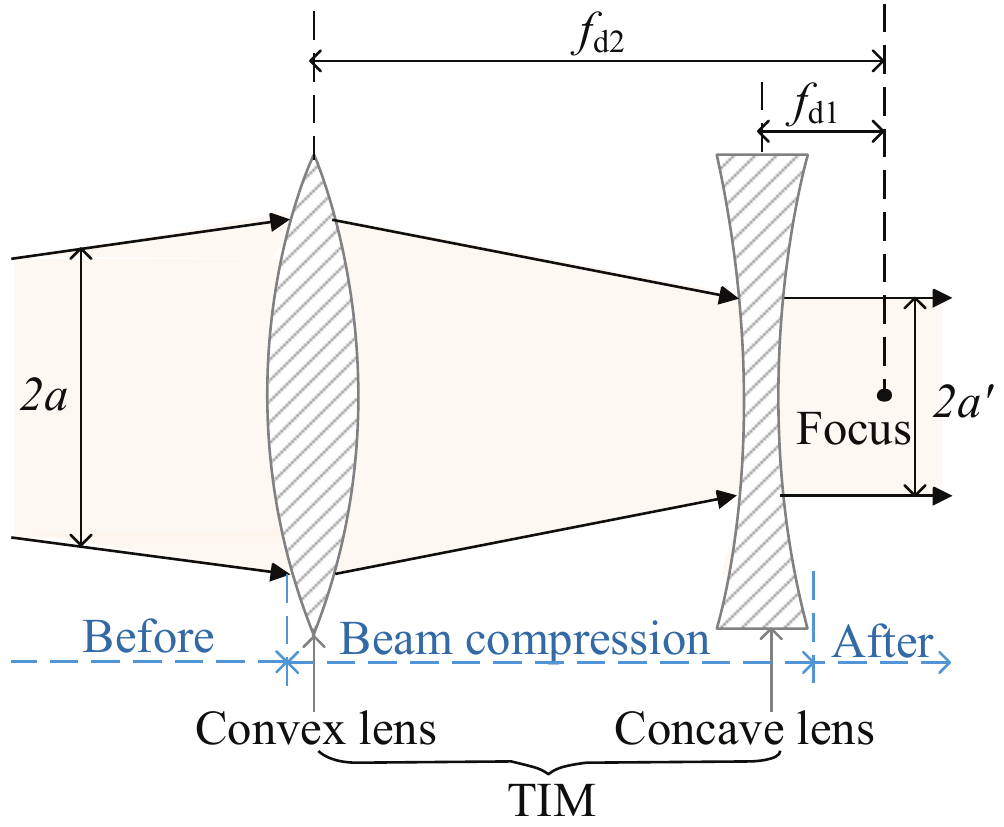}
    \caption{The beam transmission in the TIM.}
    \label{Fig_TIM}
\end{figure}

Thus, to decrease the diffraction loss during the resonant beam transmission, the diameter of resonant beam should be compressed. The role of the TIM in the system lies in compressing the resonant beam, reducing transmission losses, and improving transmission efficiency including extending transmission distance and improving output power.

As displayed in Fig.~\ref{Fig_TIM}, the TIM in the TIM-RBS comprises i) a concave lens with a focal length $f_{\rm{d1}}$, and ii) a convex lens with a focal length $f_{\rm{d2}}$, wherein the two lenses are placed parallel to each other and the focal points of them coincide \cite{van1977invention}. Then, the effective aperture of the gain medium is placed at the focal position, which allows the resonant beam with the smallest diameter to pass through the gain medium for amplification.

\subsection{End-to-End Power Transfer}\label{}
In the TIM-RBS, the energy is transformed into electricity and data through a series of conversions: i) power pumping, ii) power transfer including beam compression through the TIM and free-space transmission, iii) power output, and iv) beam power converting into electric power and communication resources.

For power pumping, the electric power from the pumping source simulates the gain medium to excite the resonant beam. Firstly, the input electric power is converted into optical power. Then, the optical power is absorbed by the gain medium for realizing the population inversion. Finally, the inverted photons are spilled by the gain medium to form resonant beam.

After the resonant beam is excited out of the gain medium, the TIM will shape its spot. When the resonant beam is transmitted forward (i.e. from the transmitter to the receiver) entering the TIM, the concave lens will perform the first phase conversion on it, so that the beam is transmitted to the focal point of the lens. Then, the beam passes through the convex lens and undergoes the second phase change. Conversely, if the resonant beam is transmitted from the receiver to the transmitter, the resonant beam will pass through the convex lens and the concave lens in the TIM successively, which causes the phase of the beam to be reversely adjusted, and the resonant beam is emitted with a smaller spot in parallel from the concave lens.

Then, the resonant beam is transmitted from the transmitter to the receiver over the air. During the free-space transmission, the resonant beam will suffer a series of losses, such as dust and other contaminants, which will affect the end-to-end transmission efficiency. Afterwards, at the receiver, a portion of the resonant beam passes through the output reflector R$2$ converting into the output beam power for charging and communication. Another part of the resonant beam is reflected to the transmitter for amplification in the gain medium.

Overall, the energy carried by the resonant beam is transmitted from the transmitter to the receiver by the above transfer stages.

\section{Transmission Model Theoretical Analysis}\label{Section3}
The charging power and the communication resource at the receiver are decided by the input power and the transmission efficiency of each transfer stage. The linear equation between the input power $P_{\rm{in}}$ and output power $P_{\rm{out}}$ is \cite{koechner2013solid}
\begin{equation}\label{E2Epower}
\begin{aligned}
    P_{\rm{out}} = (P_{\rm{in}} - P_{\rm{th}})\eta_{\rm{g}}\eta_{\rm{s}}\eta_{\rm{m}}\eta_{\rm{t}}(1-R_2),
\end{aligned}
\end{equation}
where $P_{\rm{th}}$ is the input power threshold of enabling beam oscillation, $\eta_{\rm{g}}$ and $\eta_{\rm{s}}$ denote the stimulation efficiency and the transfer factor of gain medium, $\eta_{\rm{m}}$ represents the transmission efficiency in the telescope while $\eta_{\rm{t}}$ is the resonant beam transmission efficiency over the free-space. Besides, $R_2$ is the reflectivity of the output reflector.

Among them, $P_{\rm{th}}$, $\eta_{\rm{g}}$ and $\eta_{\rm{s}}$ are the parameters relying on the hardware, such as the gain medium, the cavity, while $\eta_{\rm{m}}$ and $\eta_{\rm{t}}$ depend on the power transfer process through each optical element including TIM, input and output reflector. To explore the transmission efficiency and the output power of the TIM-RBS, we adopt the electromagnetic field propagation inside the resonant cavity to obtain the steady distribution on the optical plane.

\subsection{Field Propagation inside the Cavity}\label{}
The finite size of the transfer aperture causes diffraction loss, and diffraction plays a major role in determining the spatial distribution of the oscillation energy \cite{li1965diffraction}. In the TIM-RBS, the resonant beam will be diffracted if it passes through a series of transfer apertures, e.g. reflectors, gain medium, and TIM. Therefore, the amplitude and phase would change, and there exist the diffraction loss during power transfer.

\begin{figure}[!t]
    \centering
    \includegraphics[scale=0.52]{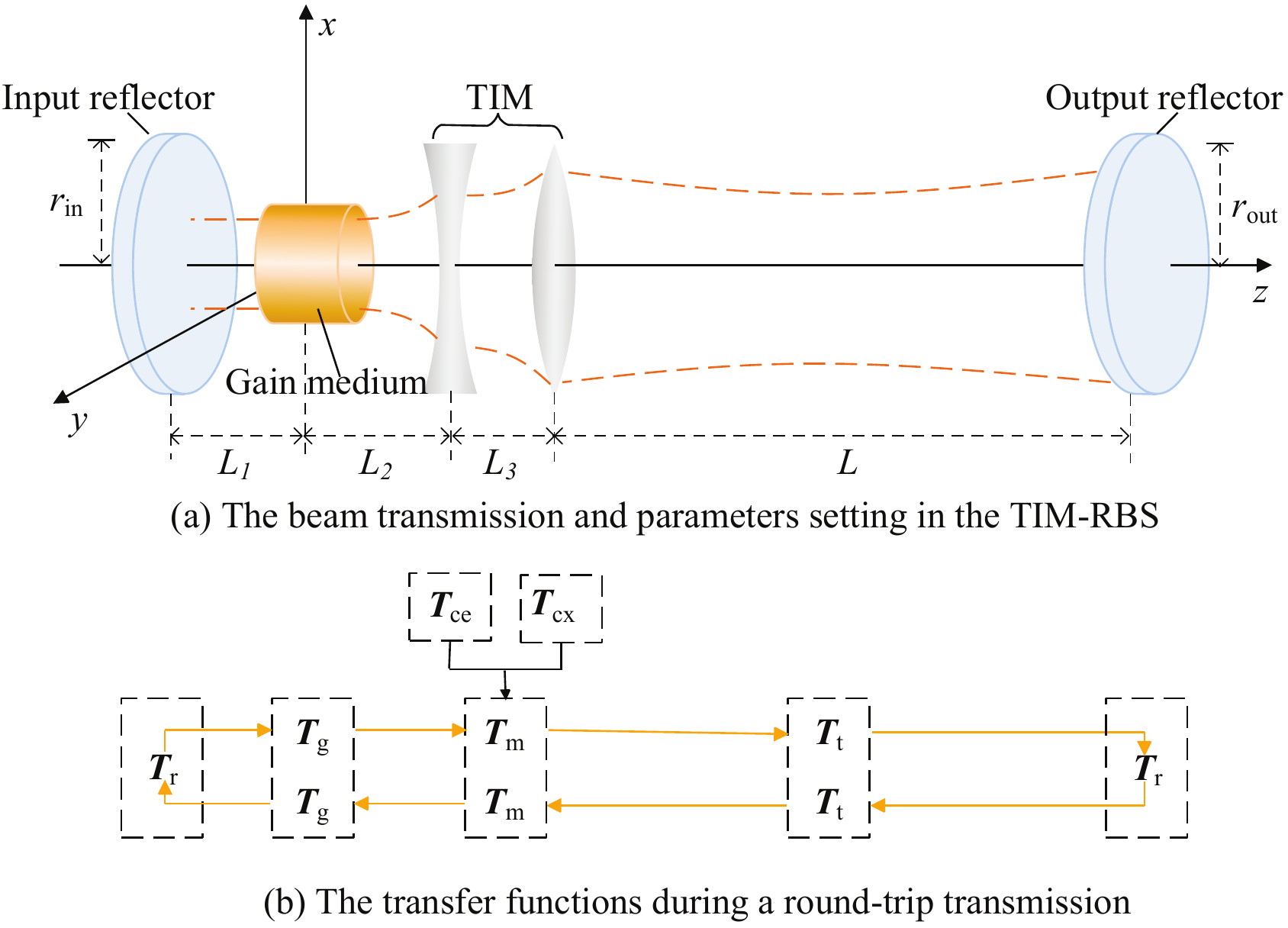}
    \caption{The parameters of optical elements and the transfer functions during a round-trip transmission in the TIM-RBS.}
    \label{Fig_Transfunction}
\end{figure}

As Fig.~\ref{Fig_Transfunction}(a) illustrated, we set the center point of the gain medium as the coordinate origin, that is $z=0$, and in which the $(x,y)$-plane is perpendicular to the $z$-axis. Then, $U(x,y,z)$ is the field mode distribution on the plane $(x,y)$ in $z$. A round-trip field propagation in the resonant cavity can be depicted in an eigenfunction as \cite{asoubar2016simulation}
\begin{equation}\label{eq:transmissionengie}
    \gamma \mathbf{U} = \mathbf{T} \mathbf{U},
\end{equation}
where the eigenvalue $\gamma$ is a complex constant related to the change in amplitude and phase. $\mathbf{U}$ represents the field distribution on a plane $U(x,y)$. $\mathbf{T}$ is the transfer function of a round-trip transmission, which can be formulated as $\mathbf{T}=\mathbf{T}_{\rm g}\mathbf{T}_{\rm m}\mathbf{T}_{\rm t}\mathbf{T}_{\rm r}\mathbf{T}_{\rm t}\mathbf{T}_{\rm m}\mathbf{T}_{\rm g}\mathbf{T}_{\rm r}$ as illustrated in Fig.~\ref{Fig_Transfunction}(b). Among them, $\mathbf{T}_{\rm g}$, $\mathbf{T}_{\rm m}$, $\mathbf{T}_{\rm t}$ and $\mathbf{T}_{\rm r}$ are the transfer function through gain medium, TIM, free space between transmitter and receiver, and reflector. Besides, $\mathbf{T}_{\rm m} = \mathbf{T}_{\rm ce}\mathbf{T}_{\rm cx}$ since the TIM comprises a concave lens and a convex lens (Fig.~\ref{Fig_Transfunction}(b)).

Using electromagnetic wave propagation theory, if the field distribution on the source wave plane $U_1(x_1, y_1, z_1)$ is known, the field distribution on the observing plane $U_2(x_2, y_2, z_2)$ can be calculated as \cite{kortz1981diffraction}
\begin{equation}\label{eq:planefield}
    U_2(x_2, y_2, z_2) = \iint_s U_1(x_1, y_1, z_1) T_{12}(x_1, y_1, x_2, y_2) ds,
\end{equation}
where $s$ is the area of the source wave plane. $T_{12}$ is the transfer function between the two planes, and it can be depicted as
\begin{equation}\label{eq:transfunction}
    T_{12} = \frac{i\exp{(-ikL)}}{\lambda L}\exp \left\{\frac{-ik}{2 L}\left[\left(x_{2}-x_{1}\right)^{2}+\left(y_{2}-y_{1}\right)^{2}\right]\right\},
\end{equation}
where $i$ is the imaginary unit, $L$ is the distance between the two planes, $\lambda$ is the wavelength of resonant beam, and $k = 2\pi/\lambda$ represents the wavenumber.

Afterwards, if the beam passes through a lens, the phase and amplitude of beam field distribution are affected and changed. Here, we treat a lens as a diffraction screen, the complex amplitude transfer function of each lens can be given by
\begin{equation}\label{eq:transplane}
\begin{aligned}
    t(x, y) &= A(x, y) P(x, y) \\
  &= A(x,y) \exp{\left[-\frac{ik}{2f}(x^2 + y^2)\right]},
\end{aligned}
\end{equation}
where $P(x, y)$ is the phase factor causing the phase changes, and $f$ is the focal length of the lens. $A(x, y)$ denotes the optical pupil function corresponding amplitude changes in the beam field, which is defined as $1$ with the area within the aperture and $0$ as the area is outside the aperture, i.e.,
\begin{equation}\label{eq:pupilfunction}
A(x, y) = \left\{
\begin{aligned}
    & 1,\ x^2 + y^2 \leq r^2 \\
    & 0,\ x^2 + y^2 > r^2
\end{aligned}
\right.
\end{equation}
where $r$ is the radius of the aperture.

\begin{figure}[!t]
    \centering
    \includegraphics[scale=0.60]{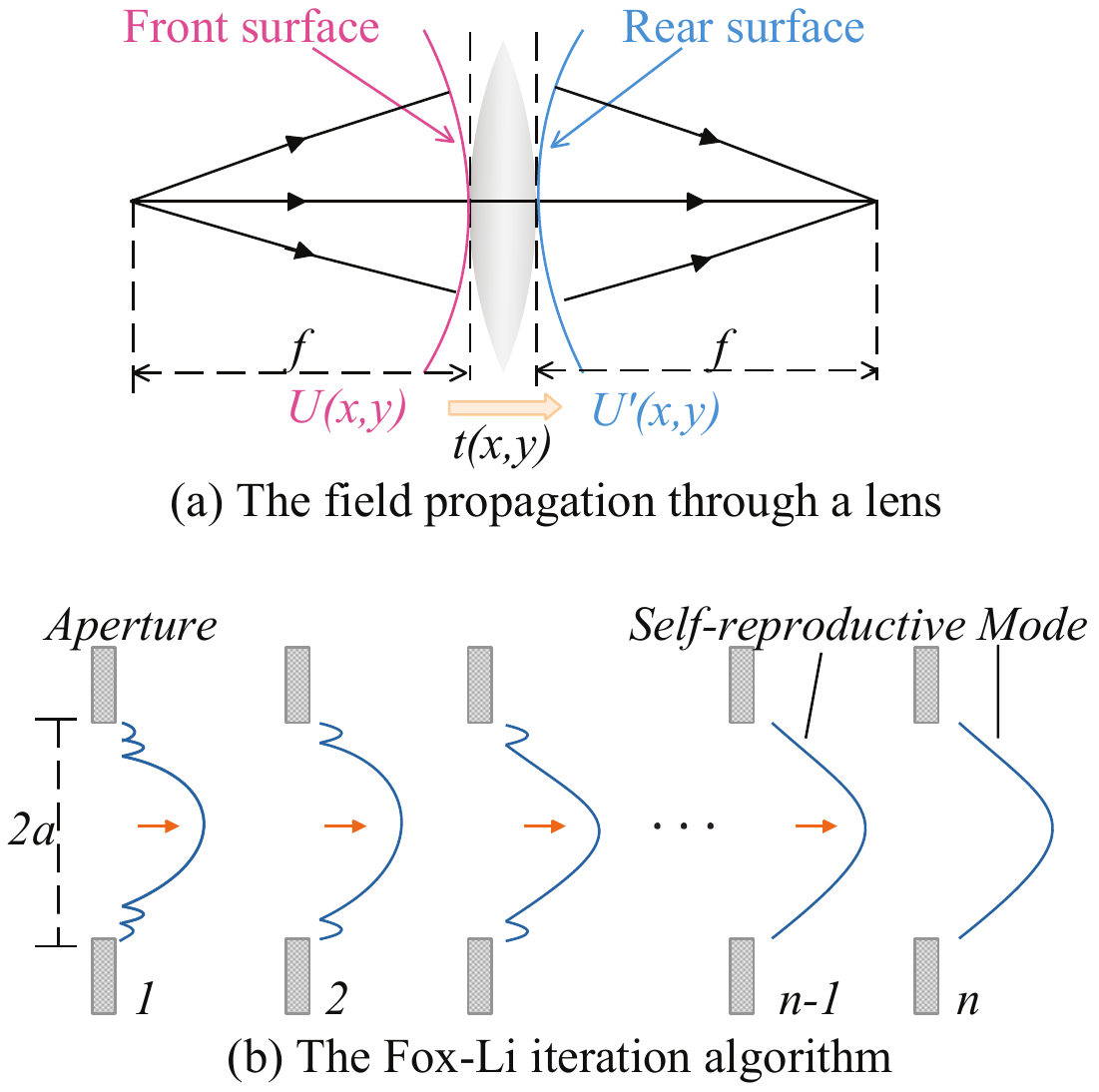}
    \caption{The field propagation through a lens from the front surface to the rear surface and the formation of self-reproductive model by the Fox-Li iteration algorithm.}
    \label{Fig_lens_fox}
\end{figure}

Then, given the transfer function of the lens, the relationship between amplitude distribution on the front surface of lens $U(x, y)$ and on the rear surface $U'(x, y)$ (Fig.~\ref{Fig_lens_fox}(a)) is
\begin{equation}\label{eq:frontandrear}
    U'(x, y) = U(x, y) t(x, y).
\end{equation}

For a round-trip power transfer in the TIM-RBS, the resonant beam passes through gain medium, TIM, output reflector, TIM, gain medium, and input reflector in sequence. To obtain the field distribution on the plane in the TIM-RBS, we adopt Fast-Fourier-Transform (FFT) method for a round-trip field propagation calculation \cite{sziklas1975mode, shen2006fast}. Based on \eqref{eq:transmissionengie}, \eqref{eq:planefield}, \eqref{eq:transfunction}, \eqref{eq:transplane}, and \eqref{eq:frontandrear}, the field propagation from the front surface $s_1$ of an optical element $E_1$ to the front surface $s_2$ of another optical element $E_2$ can be expressed as \cite{shen2006fast}
\begin{equation}\label{eq:S1s2}
\begin{aligned}
    U_2(x,y) &= \mathbf{T}_{12} \mathbf{U} \\
    &= \mathscr{F}^{-1}\left\{\mathscr{F}\{U_1(x,y) t_{1}(x,y)\} \mathscr{F}\{T_{12}(x,y)\}\right\},
\end{aligned}
\end{equation}
where $U_1(x,y)$ and $U_2(x,y)$ are the field distribution on the front surface of the optical element $E_1$ and $E_2$, $t_{1}(x,y)$ is the field transfer function from the front surface to the rear surface of $E_1$, and $T_{12}$ denotes the transfer function from $E_1$ to $E_2$ and can be depicted as
\begin{equation}\label{eq:TransfunCal2}
\begin{aligned}
    T_{12}(x,y) = \frac{j \exp{(-j k L)}}{\lambda L} \exp{ \left[\frac{-j k}{2 L}\left(x^{2}+y^{2}\right)\right]},
\end{aligned}
\end{equation}
with the distance $L$ between $E_1$ and $E_2$. Therefore, the field distribution on the input reflector after a round-trip transmission can be calculated by $\mathbf{T}_{\rm g}\mathbf{U}$, $\mathbf{T}_{\rm{ce}}\mathbf{U}$, $\mathbf{T}_{\rm{cx}}\mathbf{U}$, $\mathbf{T}_{\rm{t}}\mathbf{U}$, $\mathbf{T}_{\rm{r}}\mathbf{U}$, $\mathbf{T}_{\rm{t}}\mathbf{U}$, $\mathbf{T}_{\rm{cx}}\mathbf{U}$, $\mathbf{T}_{\rm{ce}}\mathbf{U}$, $\mathbf{T}_{\rm g}\mathbf{U}$, and $\mathbf{T}_{\rm{r}}\mathbf{U}$ in order.

Combining electromagnetic field propagation theory and FFT method, the field distribution on any plane during a round-trip transfer can be obtained. As shown in Fig.~\ref{Fig_lens_fox} (b), after several round-trip transmissions, the field distribution on a plane can achieve a stable mode, i.e., self-reproductive mode, whose amplitude and phase will not vary after the next transition. To solve \eqref{eq:transmissionengie} and obtain the self-reproductive mode, we adopt the ``Fox-Li" iteration algorithm to simulate the multiple round trips \cite{fox1961resonant}. A round-trip intra-cavity power transfer can be considered as the resonant beam passing through a series of apertures (reflector, gain medium, TIM) (Fig.~\ref{Fig_Transfunction}(a)), and one iterative process in ``Fox-Li" is equivalent to a round-trip power transfer. If a certain number of iterations are implemented, the field distribution on a plane no longer changes during transmission, the stable mode is achieved, and eventually, the self-reproductive mode is obtained (Fig.~\ref{Fig_lens_fox}(b)).

After the self-reproductive mode achieved in the TIM-RBS, the transmission loss between the two planes $s_1$, $s_2$ of the optical element $E_1$ and $E_2$ can be calculated as \cite{gordon1964equivalence}
\begin{equation}\label{eq:transmissionloss}
    \delta_{12} = \frac{\iint_{s_1}|U_1(x_1, y_1)|^2ds_1-\iint_{s_2}|U_{2}(x_{2}, y_{2})|^2ds_{2}}{\iint_{s_1}|U_1(x_1, y_1|^2ds_1},
\end{equation}
where $|U|^2$ denotes the power intensity on the plane, while the power on the plane can be obtained by double integration of power intensity $\iint|U|^2$. Afterwards, the transmission efficiency between the two planes can be depicted as
\begin{equation}\label{eq:transmissioneff}
\begin{aligned}
    \eta_{12} & = 1 - \delta_{12} \\
    & = \frac{\iint_{s_2}|U_{2}(x_{2}, y_{2})|^2ds_{2}}{\iint_{s_1}|U_1(x_1, y_1)|^2ds_1}.
\end{aligned}
\end{equation}

Hence, the transmission efficiency between any two planes can be obtained based on the electromagnetic field propagation.

\begin{figure}[!t]
    \centering
    \includegraphics[scale=0.6]{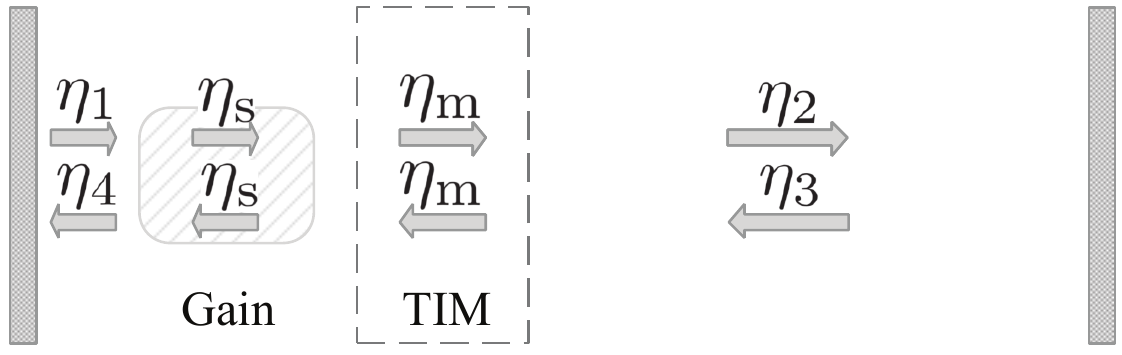}
    \caption{The transmission efficiency during a round-trip transmission.}
    \label{Fig_transeff}
\end{figure}

\subsection{End-to-End Power Transfer}\label{}
According to the output power of stable resonators, during a round-trip forward and backward traveling, the beam intensity is decreased due to diffraction losses (transfer factors $\eta_1$ - $\eta_4$), absorption inside the gain medium (transfer factor $\eta_{\rm s}$), and by output coupling (the reflectivity of the output reflector $R_2$) \cite{hodgson2005laser}. In the steady-state operation, these losses are compensated by the amplification process in the gain medium ($\eta_{\rm g}$) characterized by the small-signal gain coefficient $g_0$.

If the transfer factors (i.e. transfer efficiencies) and the small-signal gain coefficient $g_0$ are known, and the cross-sectional area of the beam is $A_{\rm b}$, the output beam power is given by \cite{hodgson2005laser}
\begin{equation}\label{eq:outputpower}
\begin{aligned}
    P_{\text {out }} &= A_{\rm b} I (1-R_2) \eta_{\rm m} \eta_2 \\
    &= A_{\rm b} I_{\rm s} \frac{(1-R_2) \eta_{\rm m} \eta_{2}}{1-R_2 \eta_{\rm m}^2 \eta_{2} \eta_{3}+\sqrt{R_2 \eta_{\rm m}^2 \eta_{\rm t}}\left[1 /\left(\eta_{1} \eta_{\rm{s}} \eta_{\rm m} \eta_{2} \right)-\eta_{\rm{s}}\right]} \\
    &\ \ \ \  \left[ g_{0} \ell-\left| \ln \sqrt{R_2 \eta_{\rm{s}}^{2} \eta_{\rm m}^2 \eta_{\rm t}} \right| \right]
\end{aligned}
\end{equation}
where $I_{\rm s}$ is the saturated beam intensity of gain medium. As shown in Fig.~\ref{Fig_transeff}, $\eta_{\rm m}$ is the transfer efficiency of the TIM, which can be calculated by multiplying the transfer efficiencies of the gain medium to the concave lens and the concave lens to the convex lens as $\eta_{\rm m} = \eta_{\rm ce} \eta_{\rm cx}$. The transmission efficiency $\eta_{\rm t} = \eta_1 \eta_2 \eta_3 \eta_4$ with $\eta_1$, $\eta_2$, $\eta_3$ and $\eta_4$ representing the forward and backward transfer efficiency between the input reflector and the gain medium, the TIM and the output reflector, respectively. Moreover, all the transmission efficiency can be calculated through \eqref{eq:transmissioneff} with the field distribution on the two planes accurately.

Additionally, based on \eqref{E2Epower} and \eqref{eq:outputpower}, to pump the resonant beam, the input power $P_{\rm{in}}$ is firstly converted into the stored power in the gain medium for stimulating the resonant beam with the efficiency $\eta_{\rm g}$, which can be expressed as \cite{hodgson2005laser}
\begin{equation}\label{eq:gaineff}
\begin{aligned}
    \eta_{\rm{g}} =\frac{AI_{\rm s}g_0\ell}{P_{\rm{in}}},
\end{aligned}
\end{equation}
where $A$ is the cross-sectional area of gain medium. Besides, to obtain the beam oscillation, the laser threshold condition should be satisfied. That is, the input power $P_{\rm{in}}$ needs to be greater than the pumping power threshold $P_{\rm{th}}$
\begin{equation}\label{eq:powerthreshold}
\begin{aligned}
    P_{\rm{th}} = \frac{A I_{\rm s} \left| \ln \sqrt{R_2 \eta_{\rm{s}}^{2} \eta_{\rm m}^2 \eta_{\rm t}} \right|}{\eta_{\rm{g}}}.
\end{aligned}
\end{equation}

Thus, in the TIM-RBS, the transmission efficiency calculation and the E$2$E power transfer analysis enable the receiver to obtain the output power with different parameters.

\subsection{Power Harvesting and Information Receiving}\label{}
At the receiver, the power splitter (PS) divides the received power into two power streams of different levels with a certain PS ratio. Afterwards, both power streams are transmitted to an energy harvester and an information decoder to achieve simultaneous power harvesting and information decoding. The information rate and the harvested energy can be balanced and optimized according to the system requirements by varying the PS ratio. Let $\theta$ denote the PS ratio for the receiver. The PV panel is responsible for converting beam power received by the receiver into electrical power, while the APD converts the light beam into electronic signals for communication.

\begin{figure}[!t]
    \centering
    \includegraphics[scale=0.6]{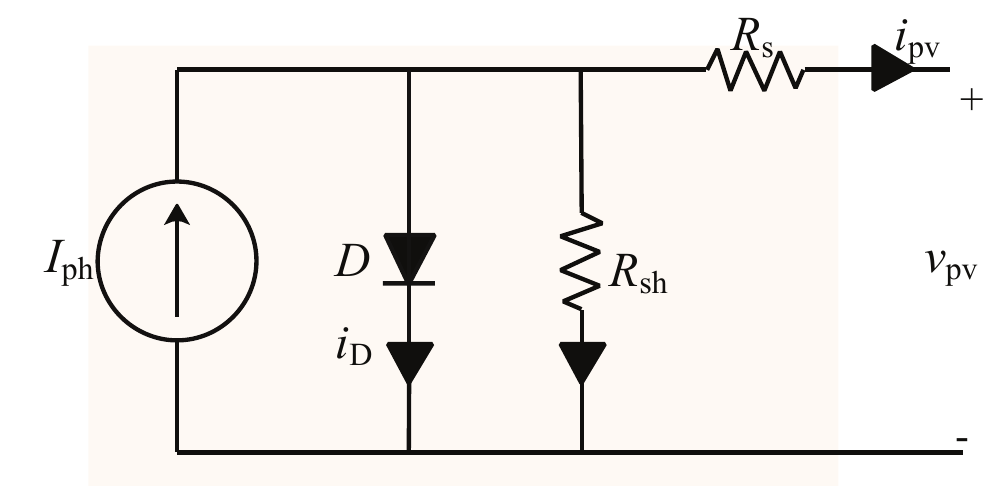}
    \caption{The equivalent circuit of a PV panel.}
    \label{Fig_pv}
\end{figure}

For power harvesting, the equivalent circuit of the single-diode model for PV panel is shown in Fig.~\ref{Fig_pv}. The meaning of the five parameters for the PV model is illustrated as follows: $I_{\rm{ph}}$ represents the photo-generated current, which depends on the input beam power of the PV panel and can be depicted as
\begin{equation}\label{current}
\begin{aligned}
    I_{\rm{ph}} = \theta \eta_{\rm{pv}} P_{\rm{out}},
\end{aligned}
\end{equation}
where $\eta_{\rm{pv}}$ is the conversion responsivity of the PV panel. $I_{\rm o}$ is the dark saturation current, $R_{\rm s}$ is the panel resistance, $R_{\rm{sh}}$ is panel parallel (shunt) resistance, and $D$ is the diode quality (ideality) factor. Afterwards, the general current-voltage characteristic of a PV panel based on the single exponential model can be expressed as \cite{sera2007pv}
\begin{equation}\label{PV}
\begin{aligned}
    i_{\rm{pv}} = I_{\rm{ph}} - I_{\rm o} \left\{\exp{\left[\frac{(v_{\rm{pv}} + i_{\rm{pv}} R_{\rm s})q}{n_{\rm s} D K T}\right]} - 1\right\} - \frac{v_{\rm{pv}} + i_{\rm{pv}} R_{\rm s}}{R_{\rm{sh}}},
\end{aligned}
\end{equation}
where $i_{\rm{pv}}$ and $v_{\rm{pv}}$ are the output current and voltage of the PV panel, and the output electric power through the PV panel can be calculated as
\begin{equation}\label{outelepower}
\begin{aligned}
      P_{\rm e} &= i_{\rm{pv}} v_{\rm{pv}} = i_{\rm{pv}}^2 R_{\rm{pv}},
\end{aligned}
\end{equation}
where $R_{\rm{pv}}$ denotes the load resistance of the PV panel. Additionally, the constants in \eqref{PV} include  Boltzmann’s constant $K$, the charge of electron $q$, the number of cells in the panel connected in series $n_{\rm s}$, and the temperature in Kelvin $T$.

For information receiving, the power stream received by APD is
\begin{equation}\label{powerapd}
\begin{aligned}
    P_{\rm{apd}} = (1 - \theta) P_{\rm{out}}.
\end{aligned}
\end{equation}
Then, the spectral efficiency, also known as information transfer rate, can be expressed as \cite{cvijetic2008performance}
\begin{equation}\label{SE}
\begin{aligned}
    C = \frac{1}{2}\log \left( 1+ \frac{S}{N}\frac{e}{2 \pi} \right),
\end{aligned}
\end{equation}
where $S$ is the signal power while $N$ is the noise power. Thereby, $S/N$ denotes the signal-to-noise ratio (SNR). Additionally, $S$ relies on the output beam power
\begin{equation}\label{SP}
\begin{aligned}
    S = \left(P_{\rm{apd}}\eta_{\rm{apd}}\right)^2,
\end{aligned}
\end{equation}
where $\eta_{\rm{apd}}$ is the photo-electric conversion efficiency of APD. $N$ is the additive Gaussian white noise (AGWN) power and can be calculated as
\begin{equation}\label{NP}
\begin{aligned}
    N & = 2q \left(P_{\rm{apd}}\eta_{\rm{apd}} + I_{\rm{bc}} \right) B_{\rm n} + \frac{4KTB_{\rm n}}{R_{\rm{apd}}},
\end{aligned}
\end{equation}
where $I_{\rm{bc}}$ is the background current, $B_{\rm n}$ is the noise bandwidth, and $R_{\rm{apd}}$ is the load resistor of the APD. The meanings of $q$, $K$, and $T$ are similar to them in \eqref{PV}.

\begin{table}[!t]
    \setlength{\abovecaptionskip}{0pt}
    \setlength{\belowcaptionskip}{-3pt}
    \centering
     \caption{Parameters related to the input/output reflectors, the TIM, and the gain medium \cite{hodgson2005laser}.}
    \begin{tabular}{ccc}
    \hline
     \textbf{Symbol} & \textbf{Parameter} & \textbf{Value}  \\
    \hline
    \bfseries{$r_{\rm{in}}$} & {Radius of the input reflector} & {$2.5\rm{mm}$} \\
    \bfseries{$r_{\rm{out}}$} & {Radius of the output reflector} & {$2.5\rm{mm}$} \\
    \bfseries{$R_1$} & {Input retro-reflector reflectivity} & {$100\rm{\%}$} \\
    \bfseries{$R_2$} & {Output retro-reflector reflectivity} & {$70\rm{\%}$} \\
    \bfseries{$L_1$} & {Length from input reflector to gain medium} & {$0.2\rm{m}$} \\
    \bfseries{$L_2$} & {Length from gain medium to TIM} & {$0.05\rm{m}$} \\
    \bfseries{$f_{\rm{d1}}$} & {Focal length of concave lens} & {$0.05\rm{m}$} \\
    \bfseries{$f_{\rm{d2}}$} & {Focal length of convex lens} & {$0.1\rm{m}$} \\
    \bfseries{$I_{\rm s}$} & {Saturation intensity} & {$1260\rm{W/cm^2}$} \\
    \bfseries{$\eta_{\rm s}$} & {Transfer coefficient of gain medium} & {$99\rm{\%}$} \\
    \bfseries{$\eta_{\rm{g}}$} & {Excitation efficiency} & {$72\rm{\%}$} \\
    \hline
    \label{table1}
    \end{tabular}
\end{table}

In summary, based on the output beam power in the TIM-RBS, the electric power and communication rate in the receiver can be obtained through \eqref{outelepower} and \eqref{SE}.

\section{Numerical Evaluation}\label{Section4}
To demonstrate the validity of the proposed transfer model in the TIM-RBS including the field propagation and the E$2$E power transfer, we will simulate the field distribution on the optical planes, the beam radius, the transmission efficiency, and the output power in this section.

\subsection{Parameters Setting}\label{}
The parameters of the TIM-RBS system are depicted in Table \ref{table1}. Among them, the input and output reflectors are plane mirrors. The Nd:YVO$_4$ is adopted as the gain medium, whose radius is less than or equal to that of the reflectors \cite{hodgson2005laser}. 

\begin{table}[!t]
    \setlength{\abovecaptionskip}{0pt}
    \setlength{\belowcaptionskip}{-3pt}
    \centering
     \caption{Parameters of energy harvesting and information receiving \cite{perales2016characterization}}.
    \begin{tabular}{ccc}
    \hline
     \textbf{Symbol} & \textbf{Parameter} & \textbf{Value}  \\
    \hline
    \bfseries{$R_{\rm{pv}}$} & {Load resistance of PV} & {$100\rm{\Omega}$} \\
    \bfseries{$\eta_{\rm{pv}}$} & {Conversion responsivity of PV} & {$0.0161\rm{A/W}$} \\
    \bfseries{$I_{\rm o}$} & {Dark saturation current} & {$9.89\times10^{-9}\rm{A}$}\\
    \bfseries{$R_{\rm s}$} & {Panel series resistance} & {$0.93\rm{\Omega}$} \\
    \bfseries{$R_{\rm{sh}}$} & {Panel parallel resistance} & {$52.6\rm{k\Omega}$} \\
    \bfseries{$D$} & {Diode quality factor} & {$1.105$} \\
    \bfseries{$n_{\rm s}$} & {Number of PV cell} & {$40$} \\
    \bfseries{$\eta_{\rm{apd}}$} & {Conversion efficiency of APD} & {$0.6\rm{A/W}$} \cite{demir2017handover} \\
    \bfseries{$q$} & {Quantity of electric charge} & {$1.6\times10^{-19}$} \\
    \bfseries{$I_{\rm{bc}}$} & {Background current} & {$5100\rm{\mu A}$} \cite{moreira1997optical} \\
    \bfseries{$B_{\rm n}$} & {Noise bandwidth} & {$811.7\rm{MHz}$} \cite{quintana2017high} \\
    \bfseries{$R_{\rm{apd}}$} & {Load resistor} & {$10\rm{k\Omega}$} \cite{xu2011impact} \\
    \bfseries{$K$} & {Boltzmann constant} & {$1.38\times10^{-23}$} \\
    \bfseries{$T$} & {Kelvin Temperature} & {$300{\rm K}$} \\
    \hline
    \label{table2}
    \end{tabular}
\end{table}

By using the PV equivalent circuit fitting measured data of a real PV cell product, we obtain the parameters of the vertical multi-junction PV cell, which are shown in Table \ref{table2} \cite{perales2016characterization}. Additionally, the parameter of the APD $\eta_{\rm{apd}}$ is from the experimental results using APD for information decoding \cite{demir2017handover}. Furthermore, other parameters related to the information receiving including $B_{\rm n}$, $I_{\rm{bc}}$, and $R_{\rm{apd}}$ are extracted from the literature \cite{quintana2017high, moreira1997optical, xu2011impact}.

\subsection{Comparison of Field Distribution}\label{}

\begin{figure*}[htbp]
\centering
\subfigure[Gain medium]{
\begin{minipage}[t]{0.23\linewidth}
\centering
\includegraphics[width=1.8in]{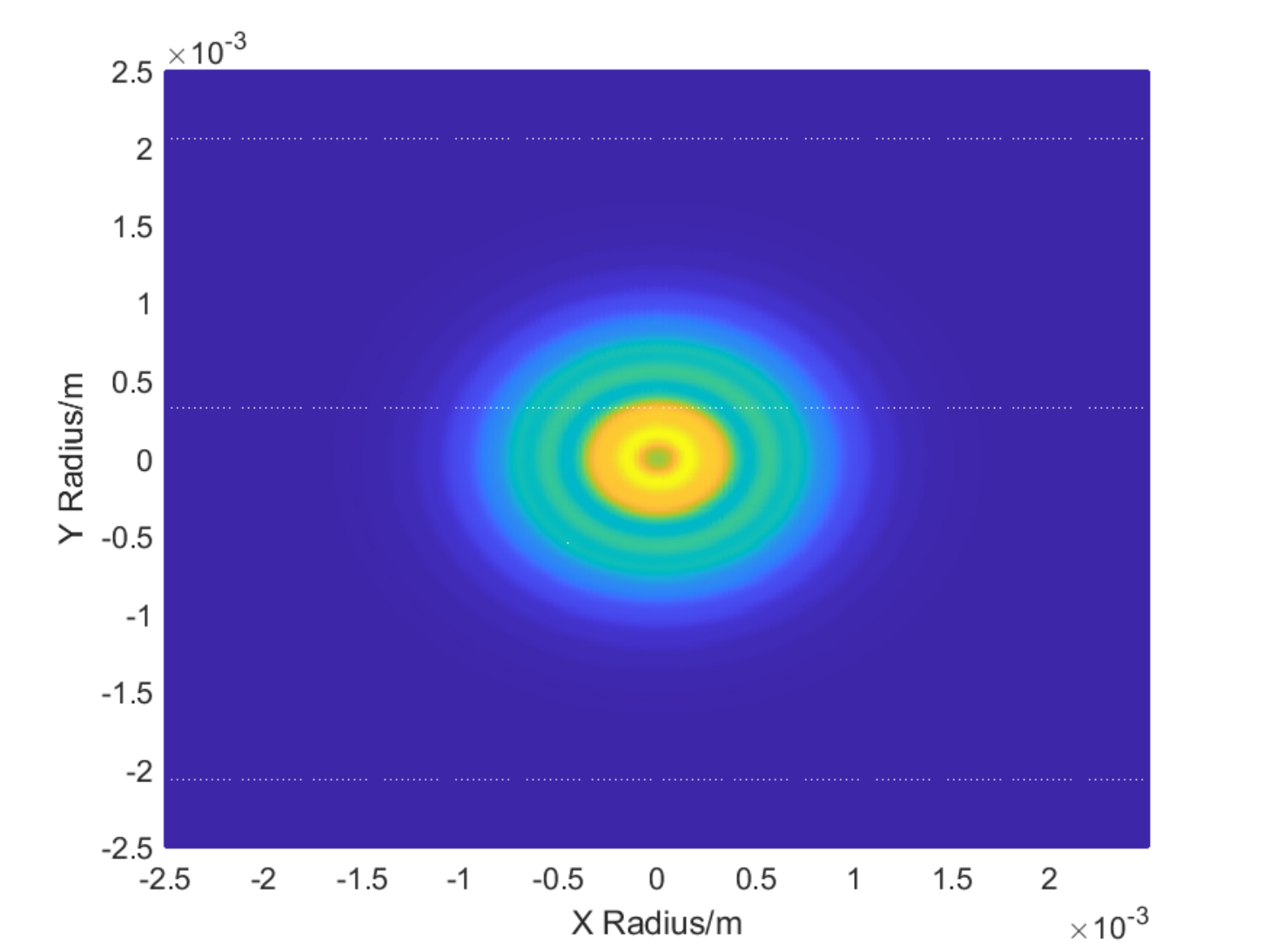}
\end{minipage}%
}%
\subfigure[Output reflector]{
\begin{minipage}[t]{0.23\linewidth}
\centering
\includegraphics[width=1.8in]{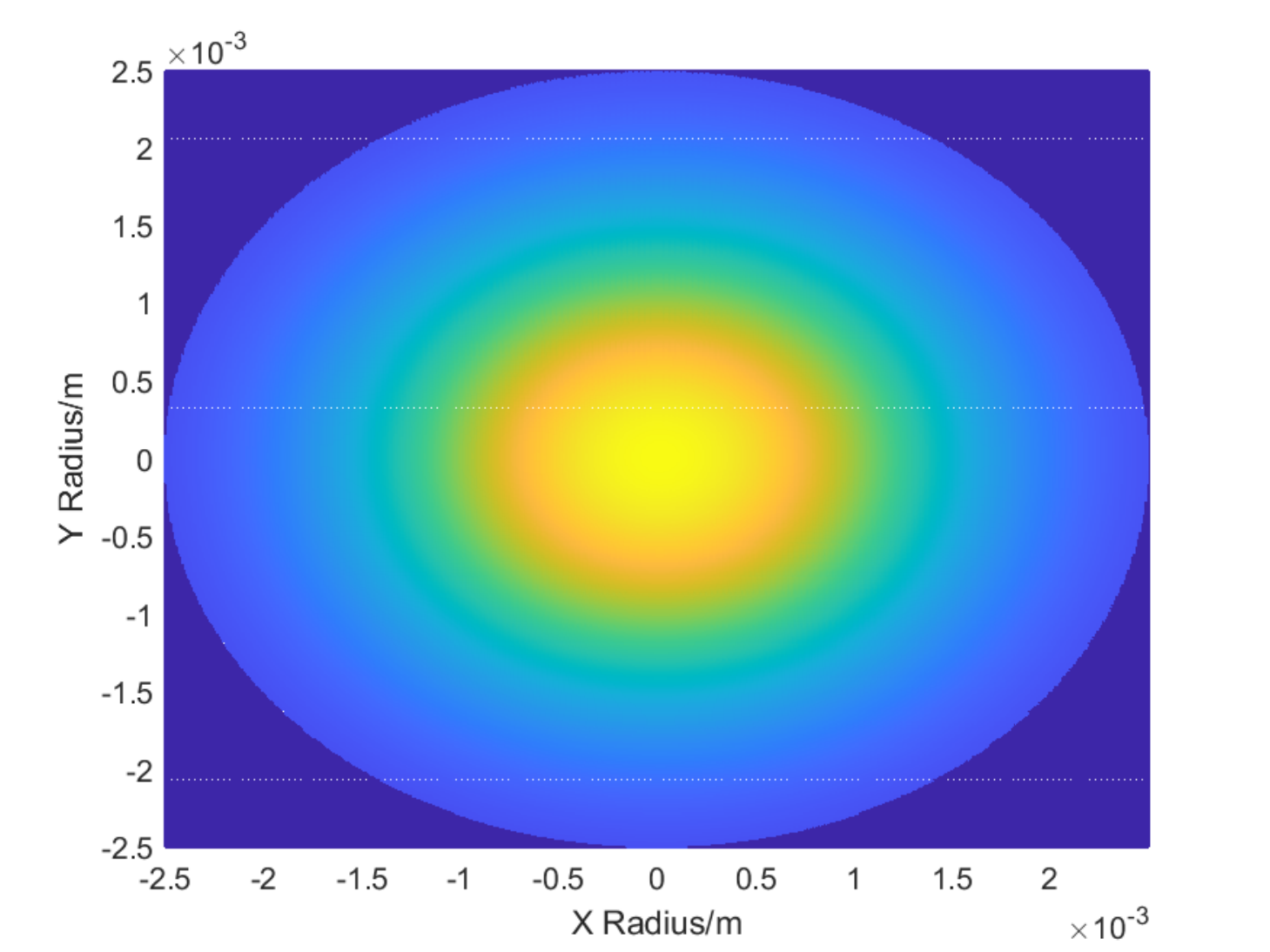}
\end{minipage}%
}%
\subfigure[Gain medium]{
\begin{minipage}[t]{0.23\linewidth}
\centering
\includegraphics[width=1.8in]{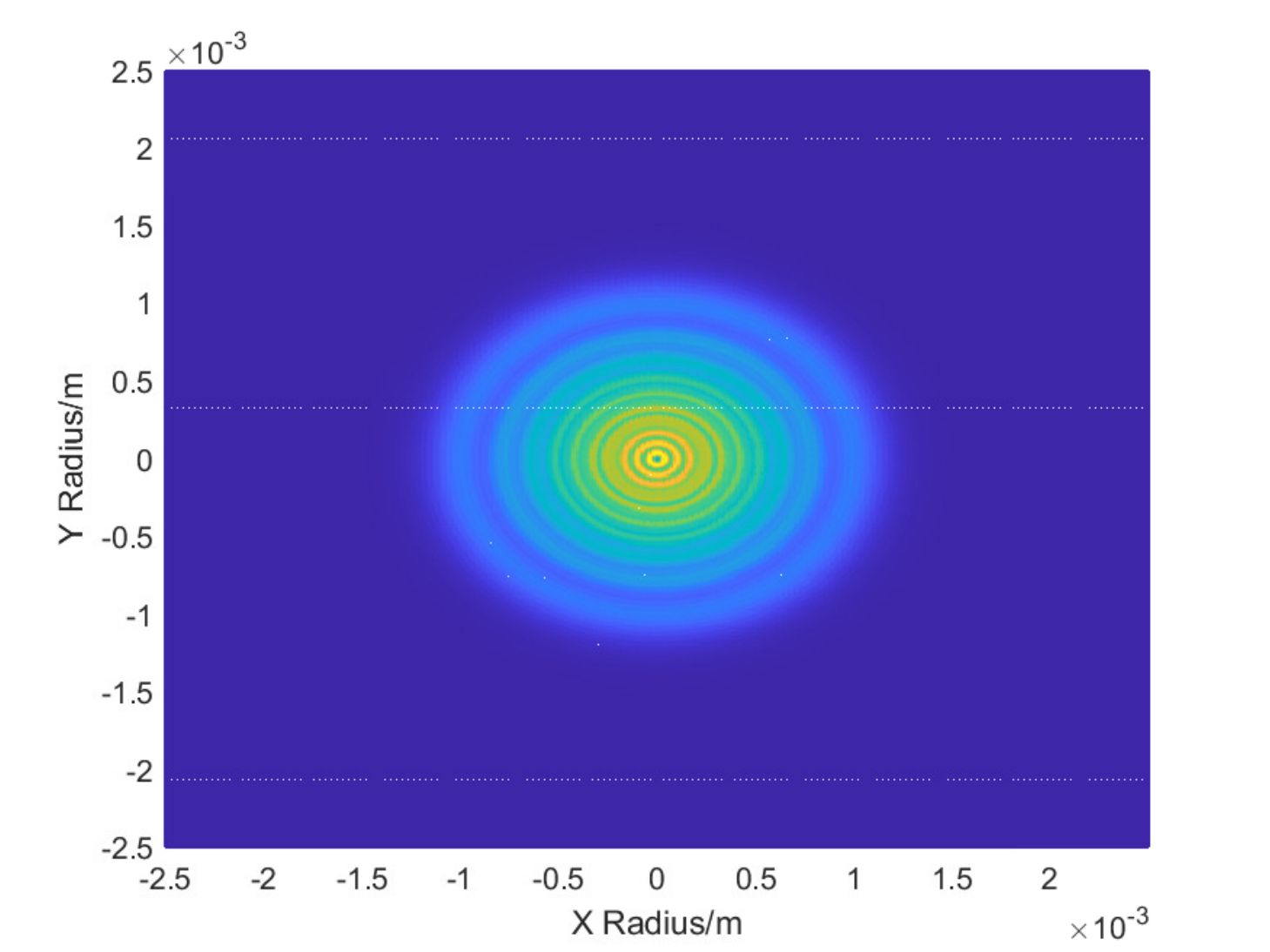}
\end{minipage}
}%
\subfigure[Input reflector]{
\begin{minipage}[t]{0.23\linewidth}
\centering
\includegraphics[width=1.8in]{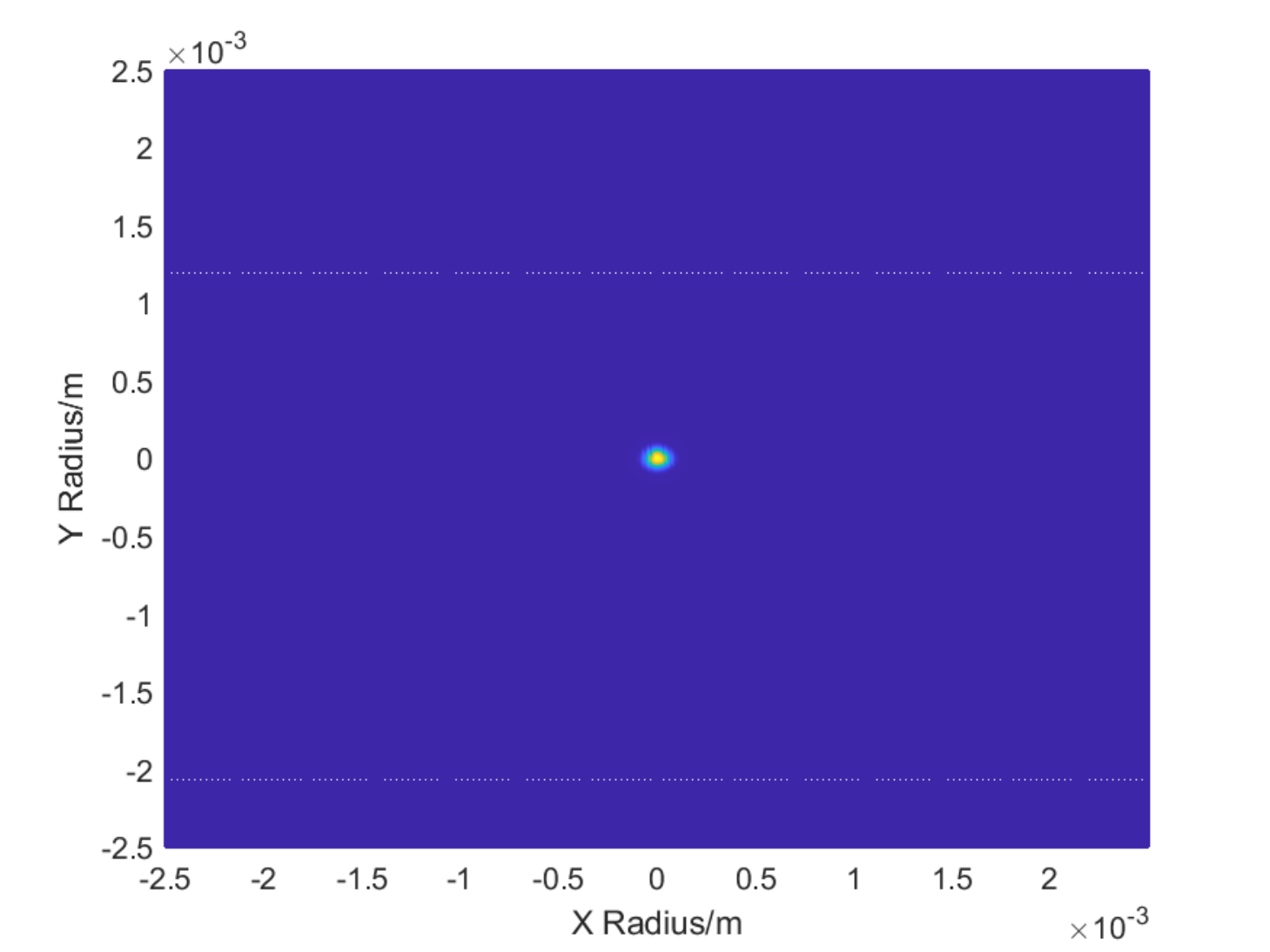}
\end{minipage}
}%
\centering
\caption{The normalized beam intensity distribution on the gain medium, output reflector, gain medium, and input reflector during a round-trip transmission in the TIM-RBS.}
\label{Fig_field_TIM}
\end{figure*}

\begin{figure*}[htbp]
\centering
\subfigure[Gain medium]{
\begin{minipage}[t]{0.23\linewidth}
\centering
\includegraphics[width=1.8in]{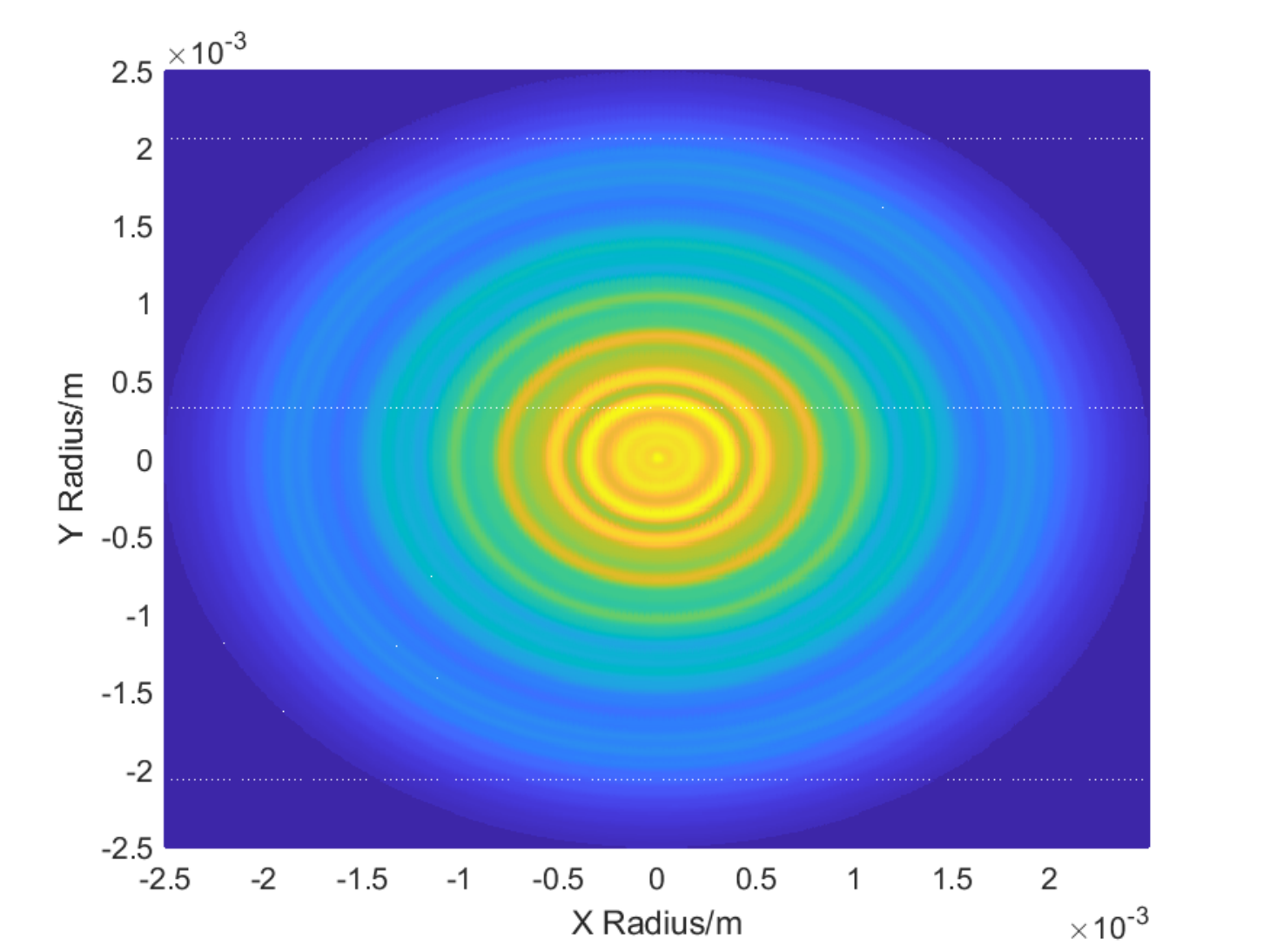}
\end{minipage}%
}%
\subfigure[Output reflector]{
\begin{minipage}[t]{0.23\linewidth}
\centering
\includegraphics[width=1.8in]{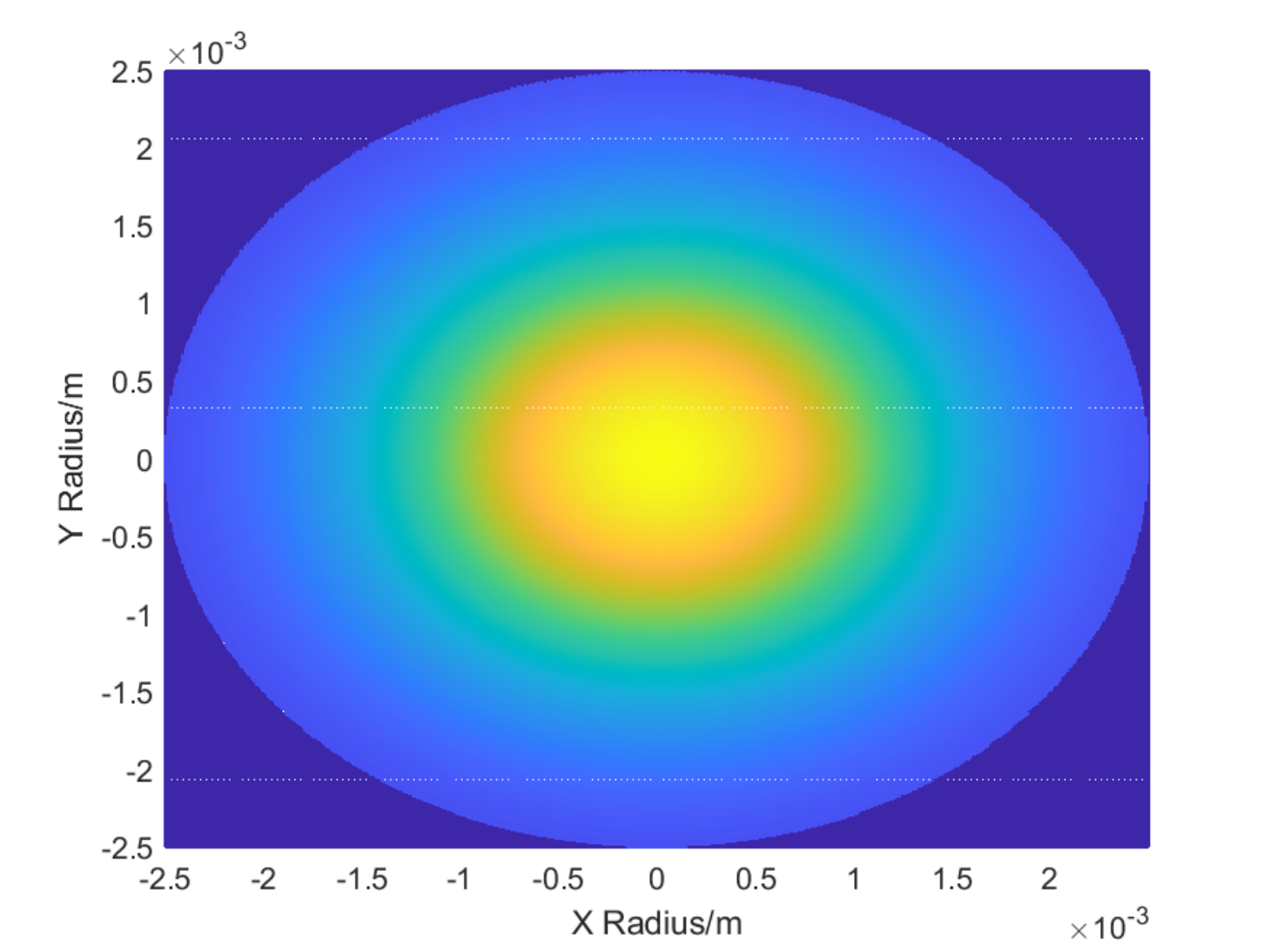}
\end{minipage}%
}%
\subfigure[Gain medium]{
\begin{minipage}[t]{0.23\linewidth}
\centering
\includegraphics[width=1.8in]{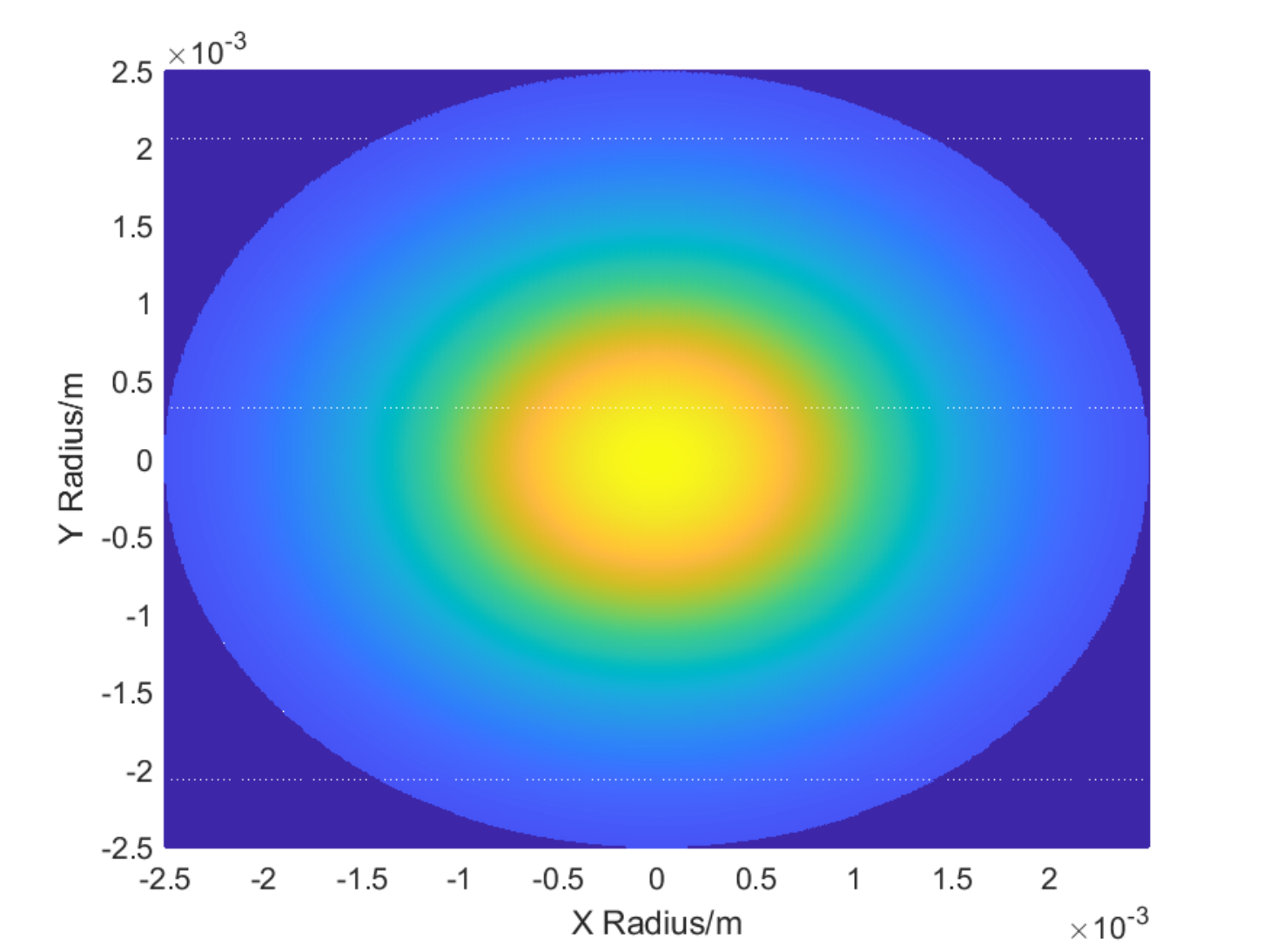}
\end{minipage}
}%
\subfigure[Input reflector]{
\begin{minipage}[t]{0.23\linewidth}
\centering
\includegraphics[width=1.8in]{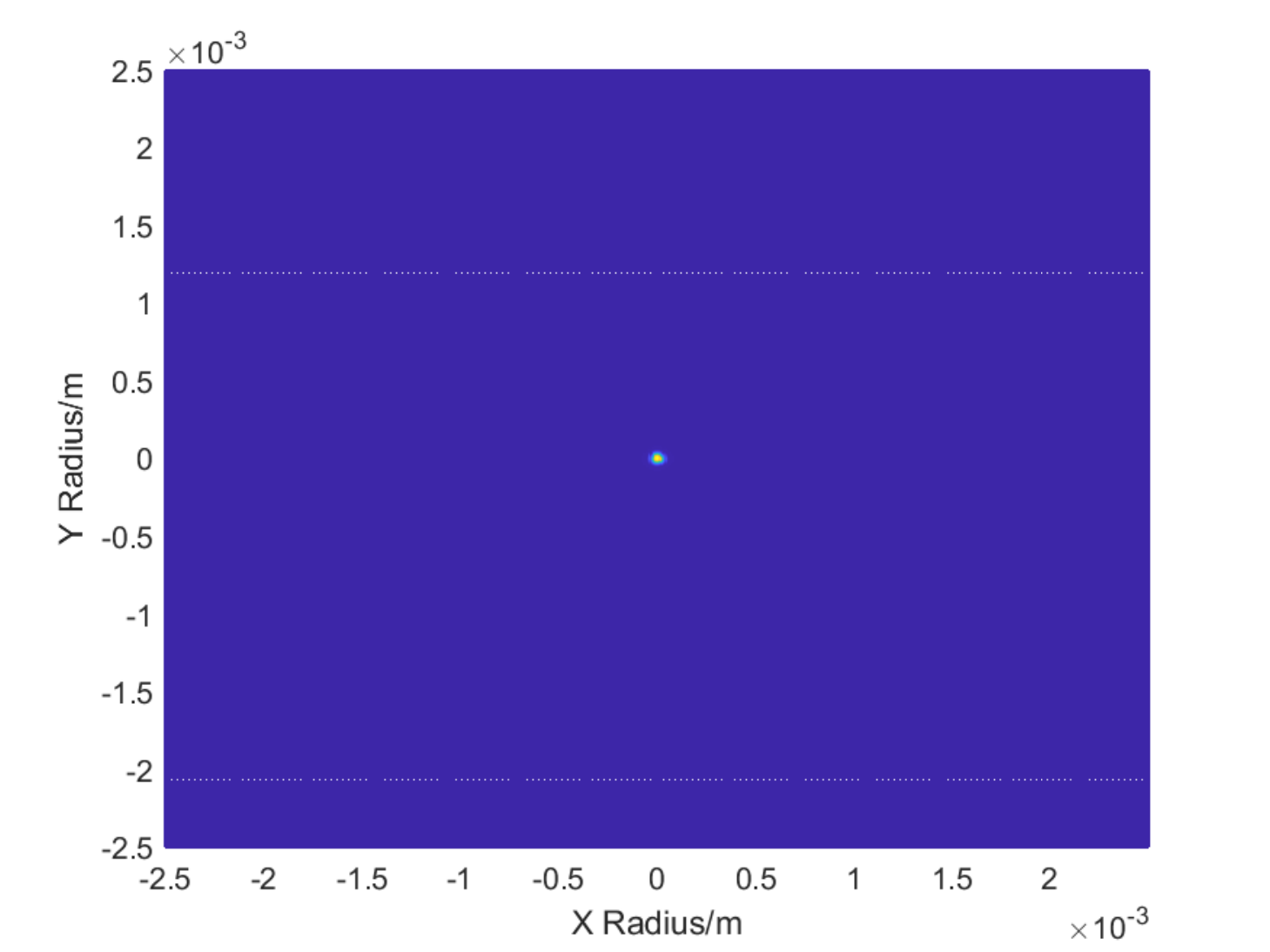}
\end{minipage}
}%
\centering
\caption{The normalized beam intensity distribution on the gain medium, output reflector, gain medium, and input reflector during a round-trip transmission in the RBS without TIM.}
\label{Fig_Field_no}
\end{figure*}

Based on \eqref{eq:planefield}-\eqref{eq:TransfunCal2}, the field distribution $U$ on the planes of the optical elements can be accurately obtained. Then, the beam intensity on the planes can be calculated by $|U|^2$. To observe the change of field distribution during a round-trip transmission and compare the compression effect of beam spot with and without TIM, we illustrate the normalized beam intensity distribution of TIM-RBS and RBS without TIM under the same radii of the input/output reflectors and gain medium with $10\rm{m}$ transmission distance in Figs.~\ref{Fig_field_TIM} and~\ref{Fig_Field_no} respectively.

During a round trip, the resonant beam passes through the gain medium, TIM, output reflector, TIM, gain medium and input reflector in turn in the TIM-RBS, while it is transmitted through the gain medium, output reflector, gain medium, and input reflector in the RBS without TIM. After the resonance in the system has achieved a steady state, that is the self-reproductive mode of resonant beam has been obtained, we study the normalized beam intensity distribution in the TIM-RBS and the RBS without TIM. 

Fig.~\ref{Fig_field_TIM}(a)-(d) shows the beam intensity distribution on the gain medium, output reflector, gain medium, and the input reflector in the TIM-RBS. From the gain medium to the output reflector (Fig.~\ref{Fig_field_TIM}(a)-(b)), the cross-sectional area of resonant beam increases because the resonant beam spreads when it travels in free space. Then, since the spot compression effect of the TIM, the cross-sectional area of the resonant beam reduces from the output reflector to the gain medium (Fig.~\ref{Fig_field_TIM}(b)-(c)). Finally, the resonant beam is focused onto the input reflector due to the function of the gain medium with convex lens properties (Fig.~\ref{Fig_field_TIM}(c)-(d)).

Similarly, Fig.~\ref{Fig_Field_no}(a)-(d) depicts the normalized beam intensity distribution on the gain medium, output reflector, gain medium, and input reflector during a round-trip transmission in the RBS without TIM. The variation of the resonant beam spot transmitting from the gain medium to the output reflector (Fig.~\ref{Fig_Field_no}(a)-(b)) and from the gain medium to the input reflector (Fig.~\ref{Fig_Field_no}(c)-(d)) is identical to that in the TIM-RBS. Differently, since there isn't the TIM between the transmitter and the receiver, the transmission from the output reflector to the gain medium without spot compression, the spot of resonant beam transmitted from the output reflector to the gain medium remains almost unchanged (Fig.~\ref{Fig_Field_no}(b)-(c)).

Therefore, the TIM in the TIM-RBS can compress the resonant beam spot during the transmission. Thus, the transmission loss in the TIM-RBS can be reduced, and finally, the transmission distance can be extended while the transmission power is able to be improved.

\subsection{Comparison of Power Transfer performance}\label{}
The TIM can suppress the transmission loss by compressing the resonant beam spot in the TIM-RBS. To illustrate the advantages of the TIM-RBS, we compare the radius of resonant beam on the gain medium (i.e. the radius at which the amplitude on the gain medium is $20\%$ of its peak value), the transmission efficiency, and the output beam power in the TIM-RBS and the RBS without TIM.

\begin{figure}[!t]
    \centering
    \includegraphics[scale=0.55]{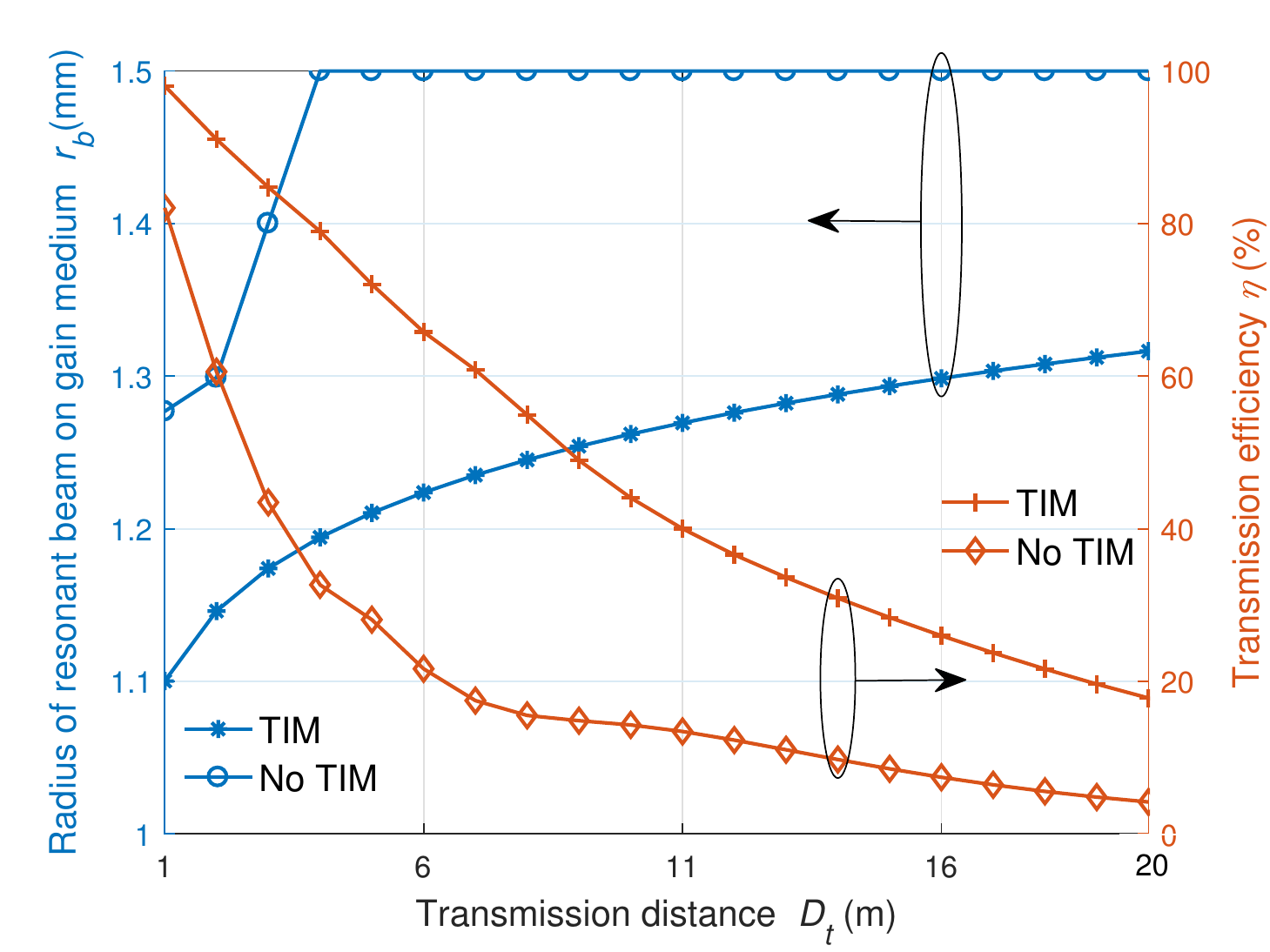}
    \caption{Comparisons of the resonant beam radius on the gain medium and the transmission efficiency in the TIM-RBS and RBS without TIM.}
    \label{Fig_TIM_no_eff_radius}
\end{figure}

Here, we analyze the power transfer performance in the TIM-RBS and RBS without TIM with $200\rm{W}$ input power $P_{\rm{in}}$ and $1.5\rm{mm}$ gain medium radius for comparison. Under the maximum transmission distance of $20\rm{m}$, the changes of resonant beam radius on gain medium $r_{\rm b}$ and transmission efficiency $\eta$ with the transmission distance $D_{\rm t}$ in the TIM-RBS and the RBS without TIM are shown in Fig.~\ref{Fig_TIM_no_eff_radius}. 

\subsubsection{For the resonant beam radius}
Firstly, due to the diffusion effect of the resonant beam, the resonant beam radius increases as the transmission distance $D_{\rm t}$ extends in the two systems. Besides, under the same transmission distance $D_{\rm t}$, the resonant beam radius on the gain medium in the TIM-RBS is less than that in the RBS without TIM. Moreover, the radius of resonant beam in RBS without TIM can increase to the same size as the gain medium if the transmission distance increases to $4\rm{m}$. For example, if $D_{\rm t}$ is $10\rm{m}$, $r_{\rm b}$ in the TIM-RBS is about $1.3\rm{mm}$ while it is about $1.5\rm{mm}$ in the RBS without TIM.

\subsubsection{For the transmission efficiency} Since the resonant beam radius increases with the extension of the transmission, the diffraction loss increases as the transmission increases. Thus, the transmission efficiency $\eta$ during a round trip (i.e., $\eta=\eta_{\rm t}\eta_{\rm m}^2$) decreases if the transmission distance grows. Moreover, the transmission efficiency in the TIM-RBS is greater than that in the RBS without TIM regardless of the transmission distance. If $D_{\rm t}$ is $10\rm{m}$, the transmission efficiency is $44\rm{\%}$ in the TIM-RBS and $14\rm{\%}$ in the RBS without TIM, respectively. 

\begin{figure}[!t]
    \centering
    \includegraphics[scale=0.55]{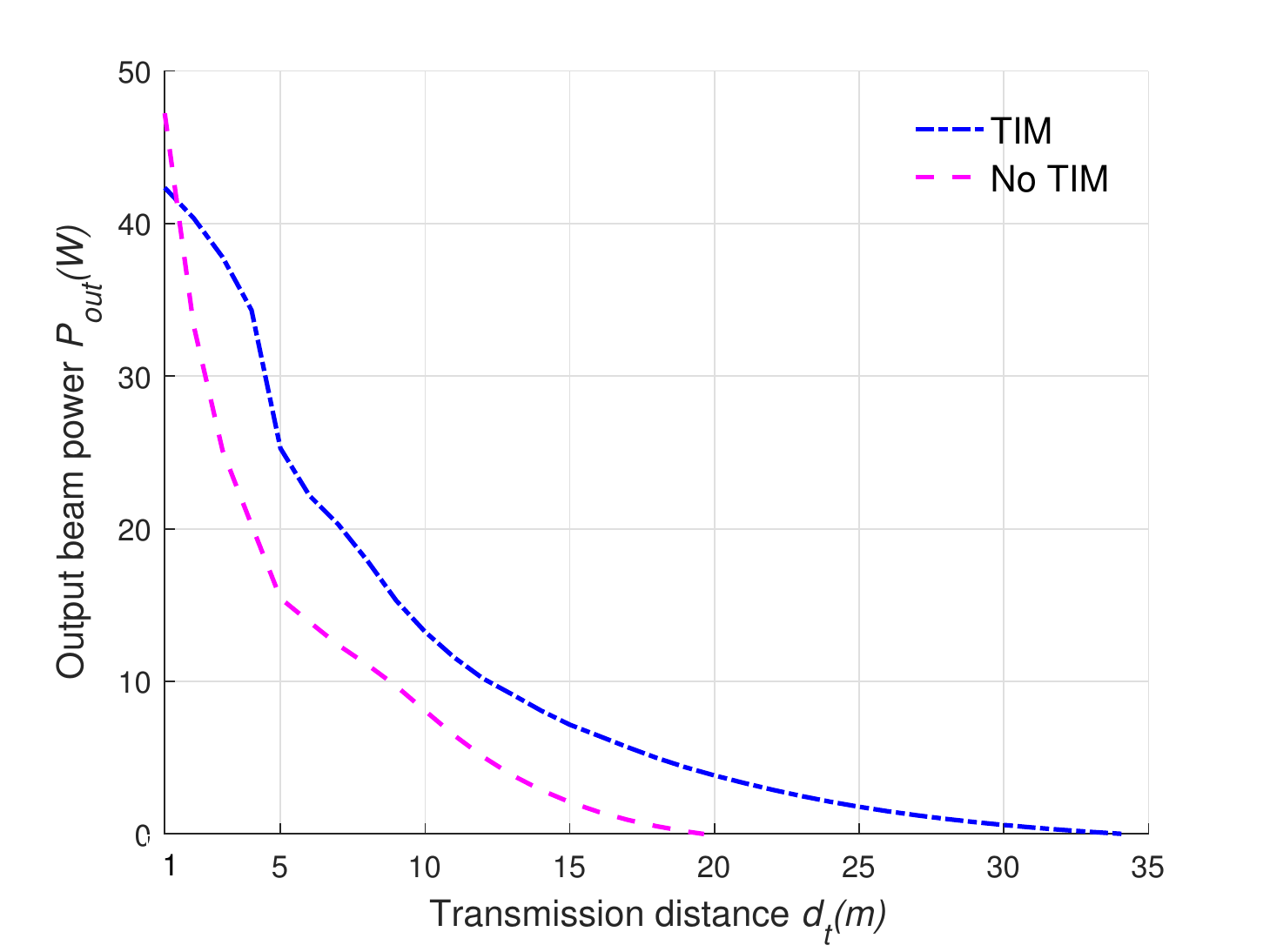}
    \caption{Comparison of the output beam power in the TIM-RBS and RBS without TIM with $200\rm{W}$ input power.}
    \label{Fig_power_TIM_no}
\end{figure}

Additionally, the variation of output beam power $P_{\rm{out}}$ of the TIM-RBS and the RBS without TIM with $200\rm{W}$ input power $P_{\rm{in}}$ is depicted in Fig.~\ref{Fig_power_TIM_no}. From Fig.~\ref{Fig_power_TIM_no}, the increase in transmission distance causes a decrease in output beam power due to the decrease of transmission efficiency. Besides, the largest transmission distance of the TIM-RBS is $35\rm{m}$ while it is $20\rm{m}$ of the RBS without TIM. 

Moreover, if the transmission distance is $1\rm{m}$, the output power in the TIM-RBS is less than that in the RBS without TIM due to the large beam spot ($A_{\rm b}=\pi r_{\rm b}^2$) and the similar transmission efficiency. Conversely, the output beam power in the TIM-RBS is greater than that in the RBS without TIM as the transmission distance is larger than $1\rm{m}$ and equal. For instance, if $D_{\rm t}$ is $10\rm{m}$, the output power is $15.30\rm{W}$ and $8.07\rm{W}$ in the TIM-RBS and the RBS without TIM, respectively.

\begin{figure}[!t]
    \centering
    \includegraphics[scale=0.55]{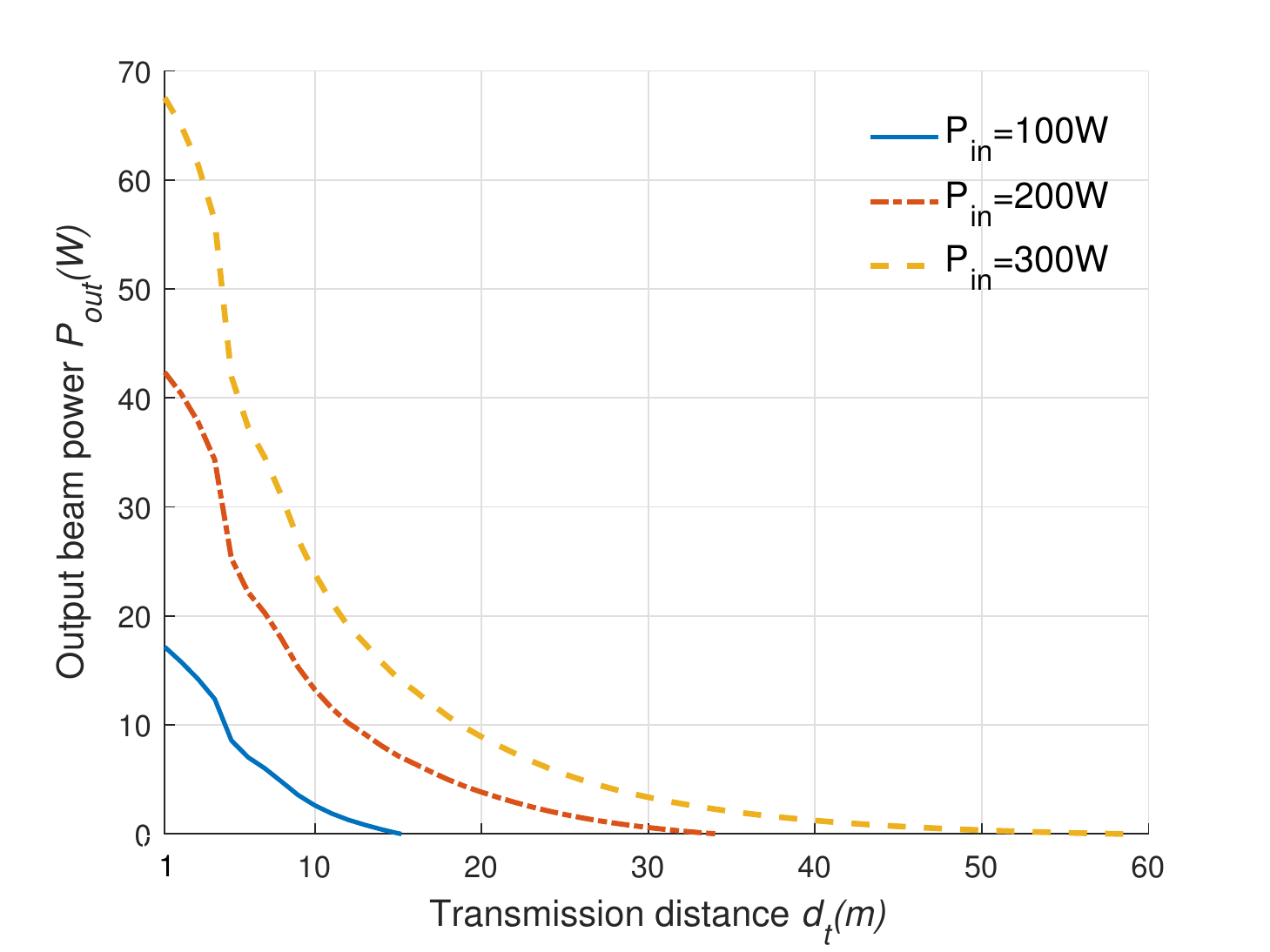}
    \caption{The changes of output beam power with different transmission distances and input power (e.g. $100\rm{W}$, $200\rm{W}$, and $300\rm{W}$) in the TIM-RBS.}
    \label{Fig_outpower_diffin}
\end{figure}

\begin{figure}[!t]
    \centering
    \includegraphics[scale=0.55]{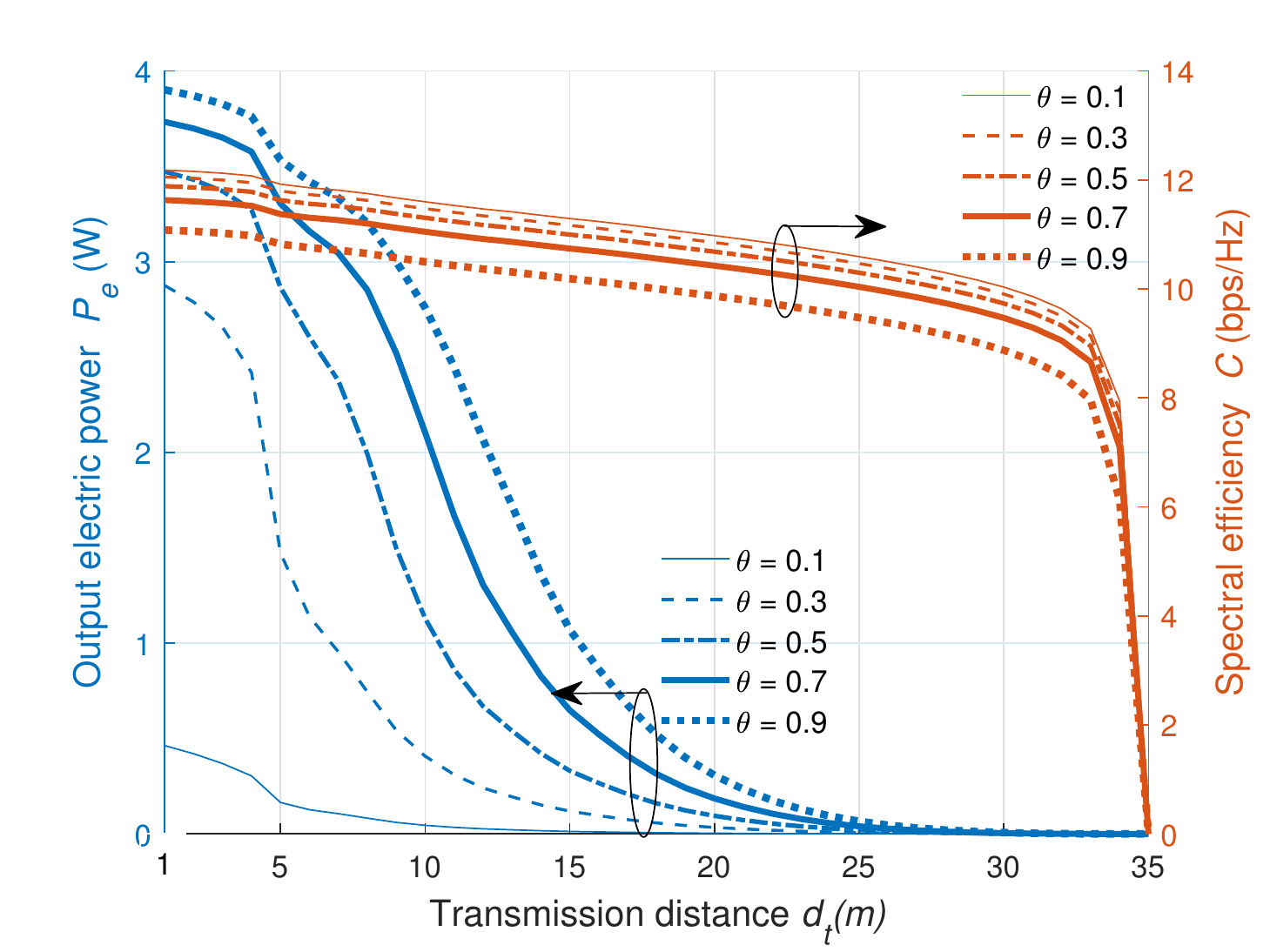}
    \caption{The changes of output electric power and spectral efficiency with different transmission distances and PS ratios under $200\rm{W}$ input power.}
    \label{Fig_comm_charging}
\end{figure}

In general, since the beam spot compression of the TIM, the diffraction loss is suppressed in the TIM-RBS. Thus, the power transfer performance in the TIM-RBS is superior to the RBS without TIM, which can be illustrated in higher output power and longer transmission distance.

\subsection{Power Transfer performance in TIM-RBS }\label{}
For the TIM-RBS, the output beam power can be obtained based on \eqref{eq:outputpower} with the determined input power $P_{\rm{in}}$. To analyze the power transfer performance of the TIM-RBS, we firstly explore the output beam power with different input power, and then study the output electric power and spectral efficiency with various PS ratios. 

Firstly, if the input power $P_{\rm{in}}$ is $100\rm{W}$, $200\rm{W}$ and $300\rm{W}$, the output beam power $P_{\rm{out}}$ obtained from \eqref{eq:outputpower} with the parameters in Table \ref{table1} is drawn in Fig.~\ref{Fig_outpower_diffin}. From Fig.~\ref{Fig_outpower_diffin}, the growth of the transmission distance $D_{\rm t}$ brings the decrease of the output beam power $P_{\rm{out}}$. If the input power is $300\rm{W}$, the output beam power is $23.87\rm{W}$, $8.91\rm{W}$, and $3.36\rm{W}$ with $10\rm{m}$, $20\rm{m}$, and $30\rm{m}$ transmission distance.

Then, the output beam power increases as the input power increases under the same transmission distance. For example, $P_{\rm{out}}$ is $2.62\rm{W}$, $13.25\rm{W}$, and $23.87\rm{W}$ if $D_{\rm t}$ is $10\rm{m}$ and $P_{\rm{in}}$ is $100\rm{W}$, $200\rm{W}$, and $300\rm{W}$. Moreover, the maximum transmission distance varies if the input power is different. That is, the maximum transmission distance extends as the input power increases. The maximum transmission distances are $15\rm{m}$, $34\rm{m}$ and $58\rm{m}$ with the $100\rm{W}$, $200\rm{W}$ and $300\rm{W}$ input power.

Additionally, the output electric power and the communication spectral efficiency (i.e. data rate) can be calculated by \eqref{outelepower} and \eqref{SE} with the known output beam power. If the input power $P_{\rm{in}}$ is $200\rm{W}$, under different transmission distance $D_{\rm t}$, the output electric power $P_{\rm e}$ and the spectral efficiency $C$ with various PS ratio $\theta$ are depicted in Fig.~\ref{Fig_comm_charging}.

Since the output beam power decreases with the extension of the transmission distance, and $P_{\rm e}$ and $C$ are proportional to the output beam power, the output electric power and spectral efficiency decrease as the transmission distance increases regardless of the PS ratio. Besides, the output electric power increases with the growth of PS ratio, while the spectral efficiency decreases as the PS ratio increases under the same transmission distance. This is because that the PS ratio determines the distribution proportion of the output beam power, i.e., if $\theta$ is $0.5$, $50\rm{\%}$ of the output beam power is converted into the output electric power and the rest is transformed into the communication resources.

Under the $10\rm{m}$ transmission distance, the output electric power is $0.41\rm{W}$ and the spectral efficiency is $11.47\rm{bps/Hz}$ with $0.3$ PS ratio while they are $2.10\rm{W}$ and $11.05\rm{bps/Hz}$ as the PS ratio is $0.7$. Moreover, the maximum electric power and spectral efficiency are about $4\rm{W}$ and $12\rm{bps/Hz}$ if the input power is $200\rm{W}$.

In summary, based on the electromagnetic field propagation and power transfer model in the TIM-RBS, the output beam power, electric power, and spectral efficiency can be calculated accurately, and they all decrease as the transmission distance extends. Moreover, the maximum transmission distance (i.e. cavity length) increases with the growth of input power.




\section{Conclusions}\label{Section5}
In this paper, we proposed an accurate transmission model based on the electromagnetic field propagation and E$2$E power transfer for TIM-RBS. Then, we evaluated the power transfer performance of the TIM-RBS by comparing it with the RBS without TIM and analyzing the output beam/electric power and spectral efficiency. Comparing the normalized beam intensity distribution and the beam radius in the TIM-RBS and RBS without TIM, we concluded that the TIM plays an important role in compressing the beam spot. Then, the comparisons of transmission efficiency and output beam power shown that the power transfer performance in TIM-RBS is better than that in the RBS without TIM.

Additionally, the power transfer performance of the TIM-RBS is related to the input power and transmission distance. The maximum transmission distance is $59\rm{m}$ with $300\rm{W}$ input power, and maximum output electric power and spectral efficiency are about $4\rm{W}$ and $12\rm{bps/Hz}$ if the input power is $200\rm{W}$. Hence, the TIM-RBS can realize longer-range SWIPT.

However, some problems in the TIM-RBS need to be solved in future research, including i) proof of safety in the system, ii) power transfer performance in the TIM-RBS with different radii of input/output reflector and gain medium, and iii) experimental validation of the TIM-RBS.

\bibliographystyle{IEEEtran}
\bibliography{references}

\begin{thebibliography}{10}
\providecommand{\url}[1]{#1}
\csname url@samestyle\endcsname
\providecommand{\newblock}{\relax}
\providecommand{\bibinfo}[2]{#2}
\providecommand{\BIBentrySTDinterwordspacing}{\spaceskip=0pt\relax}
\providecommand{\BIBentryALTinterwordstretchfactor}{4}
\providecommand{\BIBentryALTinterwordspacing}{\spaceskip=\fontdimen2\font plus
\BIBentryALTinterwordstretchfactor\fontdimen3\font minus
  \fontdimen4\font\relax}
\providecommand{\BIBforeignlanguage}[2]{{%
\expandafter\ifx\csname l@#1\endcsname\relax
\typeout{** WARNING: IEEEtran.bst: No hyphenation pattern has been}%
\typeout{** loaded for the language `#1'. Using the pattern for}%
\typeout{** the default language instead.}%
\else
\language=\csname l@#1\endcsname
\fi
#2}}
\providecommand{\BIBdecl}{\relax}
\BIBdecl

\bibitem{wu2014cognitive}
Q.~Wu, G.~Ding, Y.~Xu, S.~Feng, Z.~Du, J.~Wang, and K.~Long, ``Cognitive
  {I}nternet of {T}hings: A new paradigm beyond connection,'' \emph{IEEE
  Internet Things J.}, vol.~1, no.~2, pp. 129--143, April 2014.

\bibitem{sun2019cooperative}
L.~Sun, L.~Wan, K.~Liu, and X.~Wang, ``Cooperative-evolution-based {WPT}
  resource allocation for large-scale cognitive industrial {I}o{T},''
  \emph{IEEE Trans. Ind. Informat.}, vol.~16, no.~8, pp. 5401--5411, Aug. 2019.

\bibitem{perera2018simultaneous}
T.~D.~P. Perera, D.~N.~K. Jayakody, S.~K. Sharma, S.~Chatzinotas, and J.~Li,
  ``Simultaneous wireless information and power transfer ({SWIPT}): Recent
  advances and future challenges,'' \emph{IEEE Communications Surveys and
  Tutorials}, vol.~20, no.~1, pp. 264--302, Firstquarter 2018.

\bibitem{hou2016preliminary}
L.~Hou and S.~Tan, ``A preliminary study of thermal energy harvesting for
  industrial wireless sensor networks,'' in \emph{2016 10th International
  Conference on Sensing Technology (ICST)}, Nanjing, China, Nov. 11-13 2016,
  pp. 1--5.

\bibitem{niyato2007wireless}
D.~Niyato, E.~Hossain, M.~M. Rashid, and V.~K. Bhargava, ``Wireless sensor
  networks with energy harvesting technologies: A game-theoretic approach to
  optimal energy management,'' \emph{IEEE Wireless Communications}, vol.~14,
  no.~4, pp. 90--96, Aug. 2007.

\bibitem{krikidis2014simultaneous}
I.~Krikidis, S.~Timotheou, S.~Nikolaou, G.~Zheng, D.~W.~K. Ng, and R.~Schober,
  ``Simultaneous wireless information and power transfer in modern
  communication systems,'' \emph{IEEE Commun. Mag.}, vol.~52, no.~11, pp.
  104--110, Nov. 2014.

\bibitem{lu2015wireless}
X.~Lu, P.~Wang, D.~Niyato, D.~I. Kim, and Z.~Han, ``Wireless charging
  technologies: Fundamentals, standards, and network applications,'' \emph{IEEE
  communications surveys \& tutorials}, vol.~18, no.~2, pp. 1413--1452,
  Secondquarter 2015.

\bibitem{garnica2013wireless}
J.~Garnica, R.~A. Chinga, and J.~Lin, ``Wireless power transmission: From far
  field to near field,'' \emph{Proc. IEEE}, vol. 101, no.~6, pp. 1321--1331,
  June 2013.

\bibitem{jawad2017opportunities}
A.~M. Jawad, R.~Nordin, S.~K. Gharghan, H.~M. Jawad, and M.~Ismail,
  ``Opportunities and challenges for near-field wireless power transfer: A
  review,'' \emph{Energies}, vol.~10, no.~7, p. 1022, July 2017.

\bibitem{xia2015efficiency}
M.~Xia and S.~Aissa, ``On the efficiency of far-field wireless power
  transfer,'' \emph{IEEE Trans. Signal Process.}, vol.~63, no.~11, pp.
  2835--2847, June 2015.

\bibitem{varshney2008transporting}
L.~R. Varshney, ``Transporting information and energy simultaneously,'' in
  \emph{2008 IEEE international symposium on information theory}, Toronto, ON,
  Canada, July 6-11 2008, pp. 1612--1616.

\bibitem{ulukus2015energy}
S.~Ulukus, A.~Yener, E.~Erkip, O.~Simeone, M.~Zorzi, P.~Grover, and K.~Huang,
  ``Energy harvesting wireless communications: A review of recent advances,''
  \emph{IEEE J. Sel. Areas Commun.}, vol.~33, no.~3, pp. 360--381, Mar. 2015.

\bibitem{chen2016secrecy}
X.~Chen, D.~W.~K. Ng, and H.-H. Chen, ``Secrecy wireless information and power
  transfer: Challenges and opportunities,'' \emph{IEEE Wireless
  Communications}, vol.~23, no.~2, pp. 54--61, April 2016.

\bibitem{ku2015advances}
M.-L. Ku, W.~Li, Y.~Chen, and K.~R. Liu, ``Advances in energy harvesting
  communications: Past, present, and future challenges,'' \emph{IEEE
  Communications Surveys \& Tutorials}, vol.~18, no.~2, pp. 1384--1412,
  Secondquarter 2015.

\bibitem{fang2021end}
W.~Fang, H.~Deng, Q.~Liu, M.~Liu, M.~Xu, L.~Yang, and G.~B. Giannakis,
  ``End-to-end transmission analysis of simultaneous wireless information and
  power transfer using resonant beam,'' \emph{IEEE Trans. Signal Process.},
  vol.~69, pp. 3642--3652, June 2021.

\bibitem{liu2016dlc}
Q.~Liu, J.~Wu, P.~Xia, S.~Zhao, W.~Chen, Y.~Yang, and L.~Hanzo, ``Charging
  unplugged: {W}ill distributed laser charging for mobile wireless power
  transfer work?'' \emph{IEEE Veh. Technol. Mag.}, vol.~11, no.~4, pp. 36--45,
  Dec. 2016.

\bibitem{zhang2018distributed2}
Q.~Zhang, W.~Fang, Q.~Liu, J.~Wu, P.~Xia, and L.~Yang, ``Distributed laser
  charging: {A} wireless power transfer approach,'' \emph{IEEE Internet Things
  J.}, vol.~5, no.~5, pp. 3853--3864, Oct. 2018.

\bibitem{wang2018channel}
W.~Wang, Q.~Zhang, H.~Li, M.~Liu, X.~Liao, and Q.~Liu, ``Wireless energy
  transmission channel modeling in resonant beam charging for {I}o{T}
  devices,'' \emph{IEEE Internet Things J.}, vol.~6, no.~2, pp. 3976--3986,
  Apr. 2019.

\bibitem{hanna1981telescopic}
D.~Hanna, C.~Sawyers, and M.~Yuratich, ``Telescopic resonators for large-volume
  tem 00-mode operation,'' \emph{Optical and Quantum Electronics}, vol.~13,
  no.~6, pp. 493--507, June 1981.

\bibitem{sarkies1979stable}
P.~Sarkies, ``A stable yag resonator yielding a beam of very low divergence and
  high output energy,'' \emph{Optics Communications}, vol.~31, no.~2, pp.
  189--192, Nov. 1979.

\bibitem{li1965diffraction}
T.~Li, ``Diffraction loss and selection of modes in maser resonators with
  circular mirrors,'' \emph{The Bell System Technical Journal}, vol.~44, no.~5,
  pp. 917--932, May 1965.

\bibitem{van1977invention}
A.~Van~Helden, ``The invention of the telescope,'' \emph{Transactions of the
  American Philosophical Society}, vol.~67, no.~4, pp. 1--67, June 1977.

\bibitem{koechner2013solid}
W.~Koechner, \emph{Solid-state laser engineering}.\hskip 1em plus 0.5em minus
  0.4em\relax Springer, 2013, vol. Sixth Revised and Updated Edition.

\bibitem{asoubar2016simulation}
D.~Asoubar, ``Simulation of continuous-wave solid-state laser resonators using
  field tracing and a fully vectorial {F}ox-{L}i algorithm,'' Ph.D.
  dissertation, April 2016.

\bibitem{kortz1981diffraction}
H.~Kortz and H.~Weber, ``Diffraction losses and mode structure of equivalent
  {TEM}$_{00}$ optical resonators,'' \emph{Applied optics}, vol.~20, no.~11,
  pp. 1936--1940, June 1981.

\bibitem{sziklas1975mode}
E.~A. Sziklas and A.~E. Siegman, ``Mode calculations in unstable resonators
  with flowing saturable gain. 2: Fast fourier transform method,''
  \emph{Applied Optics}, vol.~14, no.~8, pp. 1874--1889, Aug. 1975.

\bibitem{shen2006fast}
F.~Shen and A.~Wang, ``Fast-{F}ourier-{T}ransform based numerical integration
  method for the {R}ayleigh-{S}ommerfeld diffraction formula,'' \emph{Applied
  optics}, vol.~45, no.~6, pp. 1102--1110, Feb. 2006.

\bibitem{fox1961resonant}
A.~G. Fox and T.~Li, ``Resonant modes in a maser interferometer,'' \emph{Bell
  System Technical Journal}, vol.~40, no.~2, pp. 453--488, Mar. 1961.

\bibitem{gordon1964equivalence}
J.~Gordon and H.~Kogelnik, ``Equivalence relations among spherical mirror
  optical resonators,'' \emph{Bell System Technical Journal}, vol.~43, no.~6,
  pp. 2873--2886, Nov. 1964.

\bibitem{hodgson2005laser}
N.~Hodgson and H.~Weber, \emph{Laser Resonators and Beam Propagation:
  Fundamentals, Advanced Concepts, Applications}.\hskip 1em plus 0.5em minus
  0.4em\relax Springer, 2005.

\bibitem{sera2007pv}
D.~Sera, R.~Teodorescu, and P.~Rodriguez, ``P{V} panel model based on datasheet
  values,'' in \emph{2007 IEEE International Symposium on Industrial
  Electronics}, Vigo, Spain, 4-7 June 2007, pp. 2392--2396.

\bibitem{cvijetic2008performance}
N.~Cvijetic, S.~G. Wilson, and M.~Brandtpearce, ``Performance bounds for
  free-space optical {MIMO} systems with {APD} receivers in atmospheric
  turbulence,'' \emph{IEEE J. Sel. Areas Commun.}, vol.~26, no.~3, pp. 3--12,
  Apr. 2008.

\bibitem{perales2016characterization}
M.~Perales, M.-h. Yang, C.-l. Wu, C.-w. Hsu, W.-s. Chao, K.-h. Chen, and
  T.~Zahuranec, ``Characterization of high performance silicon-based {VMJ} {PV}
  cells for laser power transmission applications,'' in \emph{High-Power Diode
  Laser Technology and Applications XIV}, vol. 9733, Mar. 2016, p. 97330U.

\bibitem{demir2017handover}
M.~S. Demir, F.~Miramirkhani, and M.~Uysal, ``Handover in {VLC} networks with
  coordinated multipoint transmission,'' in \emph{2017 IEEE International Black
  Sea Conference on Communications and Networking (BlackSeaCom)}, Istanbul,
  Turkey, Jun. 5-8 2017, pp. 1--5.

\bibitem{moreira1997optical}
A.~J. Moreira, R.~T. Valadas, and A.~de~Oliveira~Duarte, ``Optical interference
  produced by artificial light,'' \emph{Wirel. Netw.}, vol.~3, no.~2, pp.
  131--140, May 1997.

\bibitem{quintana2017high}
C.~Quintana, Q.~Wang, D.~Jakonis, X.~Piao, G.~Erry, D.~Platt, Y.~Thueux,
  A.~Gomez, G.~Faulkner, H.~Chun, M.~Salter, and D.~OBrien, ``High speed
  electro-absorption modulator for long range retroreflective free space
  optics,'' \emph{IEEE Photon. Technol. Lett.}, vol.~29, no.~9, pp. 707--710,
  Mar. 2017.

\bibitem{xu2011impact}
F.~Xu, M.-A. Khalighi, and S.~Bourennane, ``Impact of different noise sources
  on the performance of {PIN}-and {APD}-based {FSO} receivers,'' in
  \emph{Proceedings of the 11th International Conference on
  Telecommunications}, Graz, Austria, Jun. 15-17 2011, pp. 211--218.

\end{thebibliography}

\end{document}